\documentclass[a4paper,fleqn]{cas-sc}
\usepackage[numbers,sort&compress]{natbib}
\usepackage{algorithm}
\usepackage{algorithmic}
\usepackage{lineno}
\usepackage{color}
\usepackage{bm}

\def\tsc#1{\csdef{#1}{\textsc{\lowercase{#1}}\xspace}}
\tsc{WGM}
\tsc{QE}
\tsc{EP}
\tsc{PMS}
\tsc{BEC}
\tsc{DE}
\begin{document}
\let\WriteBookmarks\relax
\def\floatpagepagefraction{1}
\def\textpagefraction{.001}

\shorttitle{Theta-regularized Kriging: Modelling and Algorithms}

\shortauthors{Xuelin Xie et~al.}

\title [mode = title]{Theta-regularized Kriging: Modelling and Algorithms}
\tnotemark[1]

\tnotetext[1]{This document is the results of the research project funded by the National Key Research and Development Program of China (Grant No. 2023YFA1000103) and the National Natural Science Foundation of China (Grant No.12371424).}

\author[1]{Xuelin Xie}[orcid=0000-0003-1135-4659]
\ead{xl.xie@whu.edu.cn}

\author[1,2]{Xiliang Lu}[orcid=0000-0002-7592-5994]
\ead{xllv.math@whu.edu.cn}
\cormark[1]

\affiliation[1]{organization={School of Mathematics and Statistics},
    addressline={Wuhan University},
    city={Wuhan},
    postcode={430072},
    country={China}}
\affiliation[2]{organization={Hubei Key Laboratory of Computational Science},
    addressline={Wuhan University},
    city={Wuhan},
    postcode={430072},
    country={China}}

\cortext[cor1]{Corresponding author}

\begin{abstract}
\sep {To obtain more accurate model parameters and improve prediction accuracy, we proposed a regularized Kriging model that penalizes the hyperparameter theta in the Gaussian stochastic process, termed the Theta-regularized Kriging. We derived the optimization problem for this model from a maximum likelihood perspective. Additionally, we presented specific implementation details for the iterative process, including the regularized optimization algorithm and the geometric search cross-validation tuning algorithm. Three distinct penalty methods, Lasso, Ridge, and Elastic-net regularization, were meticulously considered. Meanwhile, the proposed Theta-regularized Kriging models were tested on nine common numerical functions and two practical engineering examples. The results demonstrate that, compared with other penalized Kriging models, the proposed model performs better in terms of accuracy and stability. The source code for the proposed method is publicly available at \url{https://github.com/xuelin-xie/Regularized-Kriging}.}
\end{abstract}

% Keywords

%%% linenumber
%\linenumbers
%\let\oldequation\equation
%\let\oldendequation\endequation
%\renewenvironment{equation}{\linenomathNonumbers\oldequation}{\oldendequation\endlinenomath}
%\let\oldalign\align
%\let\oldendalign\endalign
%\renewenvironment{align}{\linenomathNonumbers\oldalign}{\oldendalign\endlinenomath}

\maketitle
\section{Introduction}\label{sec1}
Most real-world problems usually require a large number of expensive simulations and optimization experiments to obtain an optimal solution for the target problem. For instance, optimization problems in airfoil design and vehicle crashworthiness often have high accuracy requirements \citep{bib1,bib2}. As a result, many designers use computational fluid dynamics  \citep{bib3} and finite element methods \citep{bib4} to find optimal solutions for these problems. Unfortunately, executing such high-precision models is very time-consuming and often requires hundreds or thousands of evaluations to complete. Despite the rapid growth in computer execution speed and performance, the demand for computer simulation continues to grow as society rapidly evolves. Existing computer performance is still not sufficient to meet people's needs. {To overcome this limitation, surrogate models have been developed to significantly reduce the computational time and resources required for simulation experiments.} Common surrogate models include response surface method, random forests, support vector machines, artificial neural networks, and Gaussian process regression (also known as Kriging \citep{bib7}). Recently, these methods have been widely used in various computer experiments.

{Kriging, one of the most popular interpolation models, has recently attracted substantial attention from researchers due to its flexible structure and capability to provide uncertainty estimates for predictions \citep{bib8,bib9}.} Arun compared several interpolation techniques, such as Spline curve, Inverse distance weighting, Kriging, and Nearest neighbor. His investigation showed that Kriging interpolation outperformed than the other methods in most cases \citep{bib10}. Pavlicek et al. \cite{bib11} evaluated the modeling possibilities of strongly nonlinear heating coupled problems using artificial neural networks, random forests, and Gaussian process regression. They concluded that Gaussian process regression can be preferred because it provides a very smooth function as well as information about prediction uncertainty. {Johari et al. utilized the Kriging and finite element method to investigate the stability of soil slopes \cite{bibjohari, bib57}, providing strong support for engineering design and risk management.} With the increasing demand for surrogates, the Kriging method has been applied in various fields, including geostatistics \citep{bib12}, computer science \citep{bib13}, financial engineering \citep{bib14}, and environmental remote sensing \citep{bib15}. Kriging interpolation boasts several significant advantages; however, the training of Kriging models involves inverting the correlation matrix of $n \times n$ (where $n$ is the number of training samples) and solving related optimization parameters. When the dimension of the problem is high or the number of training set samples is large, the Kriging model has considerable computational complexity \citep{bib16}. On the other hand, surrogate-based optimization methods heavily depend on the accuracy of the surrogates, posing challenges in finding an improved optimum when the global surrogate lacks accuracy \citep{bib17}. Consequently, a crucial contemporary challenge is how to effectively reduce the computational complexity and improve the accuracy of Kriging interpolation.

{Recent advances in machine learning (ML) and deep learning have attracted a lot of attention from the engineering community and have been incorporated into various. Over the past few years, numerous advanced deep-learning models have been developed due to the great success of machine learning and deep learning in various fields.} Nevertheless, concomitant with these advancements, the complexity of these models is increasing rapidly, leading to more overfitting problems \citep{bib23}. {The overfitting problem is that the model performs well on training data but poorly on test data}. In different ML algorithms, the overfitting phenomenon can be effectively addressed by regularization techniques \citep{bib24}. Regularization usually involves adding penalties to the complexity of the model and has been widely studied in various fields, such as mathematics, statistics, computer science, and engineering \citep{bib25,bib26,bib27,bib28}. Due to the complexity of interpolation, Kriging models are also more susceptible to overfitting, occasionally resulting in suboptimal performance on test sets.

In previous studies, researchers have employed regularization techniques in the engineering design optimization of Kriging models to improve generalization performance and calculation accuracy. For example,  Hung \cite{bib29} proposed a Penalized blind Kriging method (PBK) that incorporates the variable selection mechanism into the Universal Kriging (UK) model. Instead of selecting the basis functions directly, this model employs regularization techniques to shrink the coefficients of the regression basis functions. However, Hung only considered the Lasso penalty at that time. To this end, Zhang et al. \cite{bib30} extended the PBK model to Ridge and Elastic-net penalties, and proposed a regularization method to construct the blind trend function in the Kriging model. Certainly, this method essentially penalizes the regression process function of the Kriging model. {Park \cite{bib31} combined the Least Angle Regression algorithm and cross-validation to propose an improved Penalized blind Lasso Kriging method.} This method not only enhances prediction performance but also avoids the necessity for iterative calculations. {Recently, Zhao et al. \cite{bibzhao} proposed a Modified Penalized Blind Kriging (MPBK) method. They considered the elastic net penalized blind Kriging model under Cholesky decomposition and used 10-fold cross-validation to solve the regularization parameters.} However, it's crucial to note that {the Kriging method can be seen as a combination of a regression process and a stochastic process. The regression process provides an overall prediction trend, while the stochastic process affects local errors.} All the methods discussed above primarily impose penalties on the regression process of the Kriging model and do not explicitly focus on improving the stochastic process.

To some extent, the essence of penalizing the regression process is to constrain the number or weights of the basis functions by compressing the coefficients of the regression basis functions. Although this method is somewhat effective in capturing the overall prediction trend, it may encounter difficulties when dealing with complex local information in the problem. In fact, the Kriging stochastic process provides a local trend prediction of the model, and penalizing the stochastic process proves beneficial in achieving a more accurate model, especially when the local region is more complex. In addition, the coefficients and process variance of the Kriging regression process are associated with and depend entirely on the correlation parameters of the stochastic process. Regrettably, penalizing the parameters of the Kriging model stochastic process is still inadequate in theory and computer implementations, despite it being considerably important. Hence, exploring the incorporation of regularization techniques into the stochastic process of Kriging models is a significant issue worth considering.

Based on the above considerations, we proposed a regularized Kriging model that penalizes the Gaussian stochastic process parameter $\bm{\theta}$. For brevity, we denote this model as the Theta-regularized Kriging (TRK) model. Three kinds of regularization penalties were considered, namely Lasso ($l_1$ regularization), Ridge ($l_2$ regularization), and Elastic-net. {Compared with the UK, PBK, and MPBK models, the TRK model not only has certain advantages in finding the minimum value of the objective function but also has more chances to obtain an accurate parameter solution.} We derived the optimization problem for the TRK model from the viewpoint of maximizing the likelihood function. Furthermore, we presented a detailed implementation of the TRK model, including the corresponding optimization algorithm and model tuning algorithm (the geometric search cross-validation tuning algorithm, GSCV algorithm). The proposed models were tested using nine numerical functions and two real-world engineering examples. The results indicate that, in the majority of cases, the TRK model outperforms other Kriging models in terms of both accuracy and stability.

The subsequent sections of this paper are organized as follows. {Section \ref{sec2} introduces the basic principles of the Kriging model and explains the initial motivation for the TRK model. In Section \ref{sec3}, the theory and algorithm of the proposed TRK model are explored in detail.} Section \ref{sec4} employs common benchmark functions and practical engineering examples to numerically validate the proposed model. Finally, in Section \ref{sec5}, we provide a summary and discussion of the results presented in this article.

\section{{Preliminaries}}\label{sec2}
\subsection{{Basics of Kriging Model}}\label{subsec1}
\begin{figure}[h]
\centering
\includegraphics[width=0.6\textwidth]{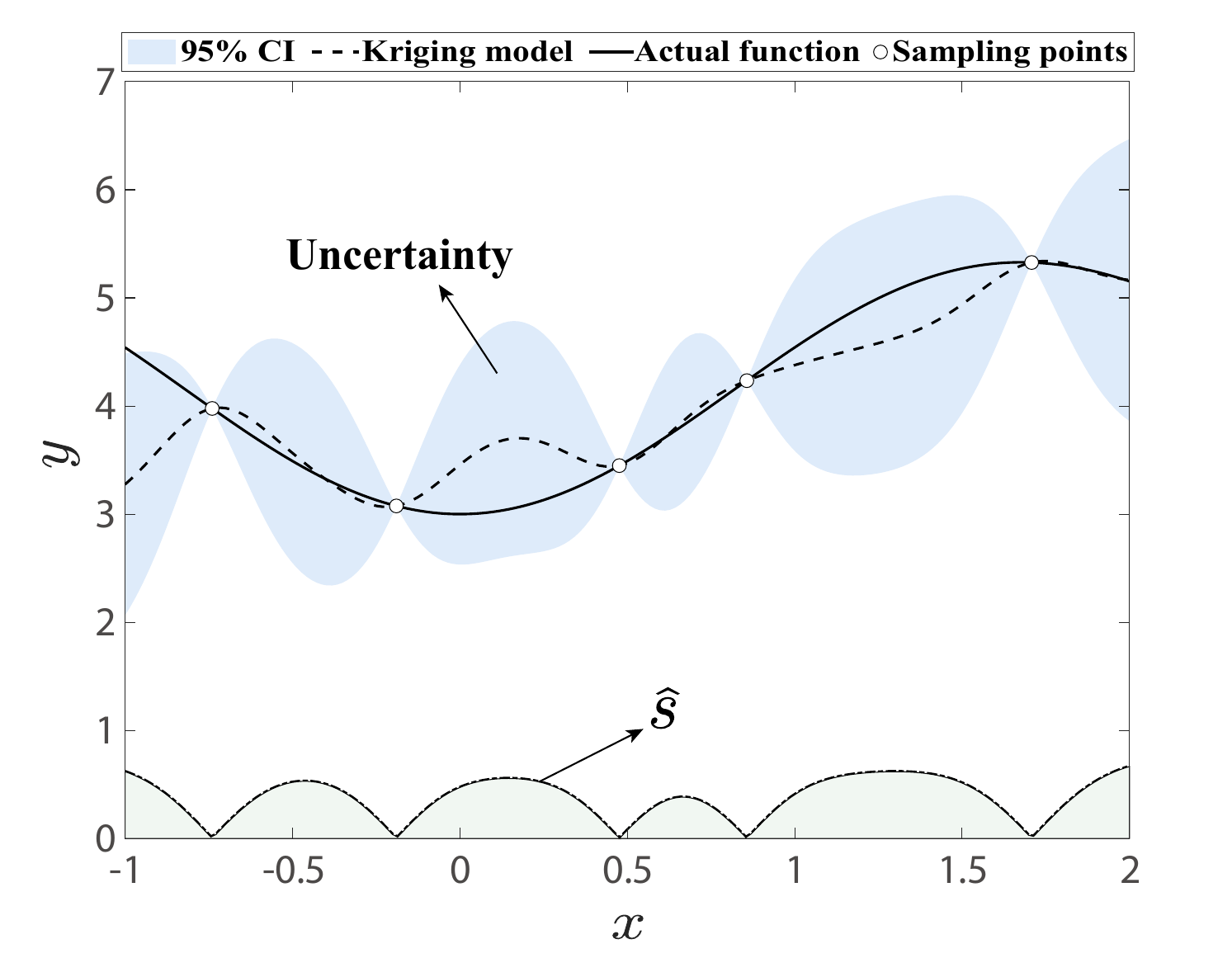}
\caption{Uncertainty estimation for Kriging prediction.}\label{fig1}
\end{figure}

The idea of the Kriging model first emerged in the field of geostatistics \cite{bib32}, and at that time, it was primarily employed for mineral prediction. Matheron extended this idea to the field of mathematical statistics and proposed the theory of regional variables \cite{bib33}. Shortly thereafter, Sacks and Lophaven et al. combined the Kriging model with computer-based design and analysis experiments \citep{bib34,bib35}. Since then, the Kriging model has received much attention from the scientific and engineering communities. The Kriging surrogate model can accurately and efficiently deal with highly nonlinear and multimodal problems. Besides, it can also provide uncertainty estimates for predictions, as illustrated in Fig. \ref{fig1}. In recent years, Kriging models have been widely used in various fields such as design optimization, modelling approximation, and structural probabilistic analysis \citep{bib36,bib37,bib38}.

The Kriging model is defined as a weighted linear combination of known points \citep{bib34}:
{\begin{equation}
\mathbf{\hat{y}}\left(\mathbf{x}\right)=\mathop{\sum }_{i=1}^{n}{{\omega }_{i}}(\mathbf{x}){\mathbf{y}_{i}}=\omega {{(\mathbf{x})}^{T}}\mathbf{y},  \label{eq1}
\end{equation} 
where ${{\omega }_{i}}(\mathbf{x})$ is the weight coefficient of ${\mathbf{y}_{i}}$, $\mathbf{y}={{\left( {\mathbf{y}_{1}},{\mathbf{y}_{2}},\ldots ,\mathbf{y}_{n} \right)}^{T}}$ is the response value matrix, and $\mathbf{\omega(x)}={{[{{\omega }_{1}}(\mathbf{x}),{{\omega }_{2}}(\mathbf{x}),...,{{\omega }_{n}}(\mathbf{x})]}^{T}}$ is the weighting vector.}

To calculate the weighting coefficients, the Kriging model assumes that the input $\mathbf{x}\in {\mathbb{R}^{D}}$ ($D$ is the problem dimension) and the corresponding output response $\mathbf{y(x)}$ have the following relationship \citep{bib39}:
\begin{equation}
\mathbf{y}\left(\mathbf{x}\right)=\sum\nolimits_{i=1}^{p}{f_{i}}(\mathbf{x}){\beta_i}+z(\mathbf{x})=f(\mathbf{x})^{T}\bm{\beta} +z(\mathbf{x}). \label{eq2}
\end{equation}
{where ${f_{i}}$ is the polynomial regression function, ${\beta_i}$ represents the coefficient of the regression function.}

Equation (\ref{eq2}) can be divided into two parts: the first part represents the linear (or nonlinear) regression process on the data, while the second part embodies the realization of the stochastic process. Concerning the regression process, $p$ denotes the number of regression basis functions, $f(\mathbf{x})$ and $\bm{\beta}$ represent function vectors and unknown coefficients, respectively. On the other hand, for the stochastic process, $z(\mathbf{x})$ signifies a Gaussian random field with a parameter space \citep{bib36}. This stochastic process assumes that when the coefficients $\bm{\beta}$ are selected appropriately, the values behave akin to "white noise."

Generally, $z(\mathbf{x})$ is assumed to follow a Gaussian process with mean 0 and variance $\bm{\sigma^2}$. Thus, for any two points ${\mathbf{x}_{i},\mathbf{x}_{j}}\in {\mathbb{R}^{D}}$, their covariance is
\begin{equation}
Cov[z(\mathbf{x}_{i}), z(\mathbf{x}_{j})]={\bm{{\sigma}^{2}}}\mathcal{R}(\bm{\theta} ,\mathbf{x}_{i},\mathbf{x}_{j}), \label{eq3}
\end{equation} 
where $i, j \in 1,2,...,n$, $n$ is the number of samples, $\bm{\theta} =({{\bm{\theta}}^{(1)}},{{\bm{\theta}}^{(2)}},...,{{\bm{\theta}}^{(D)}})$ is the hyperparameter vector of the Kriging model, and $\mathcal{R}(\bm{\theta} ,\mathbf{x}_{i},{\mathbf{x}_{j}})$ is the correlation function. Typically, the correlation function of Kriging model is assumed to satisfy the following correlation rule \citep{bib40}:
\begin{equation}
\mathcal{R}(\bm{\theta} ,\mathbf{x}_{i},\mathbf{x}_{j})=exp\left\{ -\underset{k=1}{\overset{D}{\mathop \sum }}\,{{\bm{\theta}}^{(k)}}|\mathbf{x}_{i}^{(k)}-\mathbf{x}_{j}^{(k)}{{|}^{2}} \right\}:={R_{ij}}.  \label{eq4}
\end{equation}

For each input sample $\mathbf{x}_{1},...,\mathbf{x}_{n}\in{\mathbb{R}^{D}}$, the distribution of $\mathbf{z}(\mathbf{x}_{1}),..., \mathbf{z}(\mathbf{x}_{n})$ is multivariate normal. Therefore, the Kriging model correlation matrix $\mathbf{R}$ is given as
\begin{equation}
\mathbf{R}=\left[ \begin{matrix}
\mathcal{R}(\bm{\theta} ,{\mathbf{x}_{1}},{\mathbf{x}_{1}}) & \mathcal{R}(\bm{\theta} ,{\mathbf{x}_{1}},{\mathbf{x}_{2}}) & \cdots & \mathcal{R}(\bm{\theta} ,{\mathbf{x}_{1}},{\mathbf{x}_{n}}) \\
\mathcal{R}(\bm{\theta} ,{\mathbf{x}_{2}},{\mathbf{x}_{1}}) & \mathcal{R}(\bm{\theta} ,{\mathbf{x}_{2}},{\mathbf{x}_{2}}) & \cdots & \mathcal{R}(\bm{\theta} ,{\mathbf{x}_{2}},{\mathbf{x}_{n}}) \\
\vdots & \vdots & \ddots & \vdots \\
\mathcal{R}(\bm{\theta} ,{\mathbf{x}_{n}},{\mathbf{x}_{1}}) & \mathcal{R}(\bm{\theta} ,{\mathbf{x}_{n}},{\mathbf{x}_{2}}) & \cdots & \mathcal{R}(\bm{\theta} ,{\mathbf{x}_{n}},{\mathbf{x}_{n}}) \\
\end{matrix} \right]. \label{eq5}
\end{equation} 

In addition, the vector $\mathbf{r(x)}$ is employed to represent the correlation between unknown point $\mathbf{x}$ and $n$ known sampling points:
\begin{equation}
\mathbf{r(x)}={{\left[ \mathcal{R}\left(\bm{\theta} ,\mathbf{x},\mathbf{x}_{1} \right),~\ldots ,~\mathcal{R}\left(\bm{\theta} ,\mathbf{x},{\mathbf{x}_{n}} \right) \right]}^{T}}. \label{eq6}
\end{equation}

Based on the above assumptions, the purpose of the Kriging model is to minimize the Mean Square Error (MSE) loss \citep{bib41}, {and after a series of derivations (see references \citep{bib34}), the following final expression can be obtained:}
\begin{equation}
\mathbf{\hat{y}}\left( \mathbf{x} \right)=f{{\left( \mathbf{x} \right)}^{T}}\hat{\bm{\beta}}+ {r(\mathbf{x})}^T{\gamma}^{*}, \label{eq11}
\end{equation}
where $\mathbf{\hat{\bm{\beta}}={{\left( {{F}^{T}}{R}^{-1}F \right)}^{-1}}{{F}^{T}}{R}^{-1}y}$ is the generalized least squares estimate of $\bm{\beta}$, and $\mathbf{{\gamma}^{*}={{R}^{-1}}\left( y-F\hat{\bm{\beta}} \right)}$, which is related only to the known sample points. Although the model has an explicit expression of the equations, as it is essentially non-parametric (i.e., the parameters are not fixed, and their magnitude is related to the trained data), we need to solve for the unknown parameters $\bm{\beta}$, $\bm{\sigma^2}$ and $\bm{\theta}$ of this model.

To estimate the parameters of the Kriging model, the likelihood function for the unknown parameters $\bm{\beta}$ , $\bm{\sigma^2}$, and $\bm{\theta}$ can be expressed as \citep{bib29,bib31}
\begin{equation}
L(\bm{\beta},\bm{\sigma^2},\bm{\theta}|\mathbf{y})\!=\! \!\frac{1}{{{( 2\pi\bm{{\sigma}^{2}})}^{\frac{n}{2}}}{{|\mathbf{R}|}^{\frac{1}{2}}}}\!\text{exp}\left(\!-\!\frac{\mathbf{{{(y\!-\!F\bm{\beta})}^{T}}{{R}}^{-1}(y\!-\!F\bm{\beta})}}{2{\bm{{\sigma}^{2}}}}\!\right). \label{eq12}
\end{equation}

{The solution of the Kriging model essentially involves a nested constrained optimization problem for the parameters $\bm{\beta}$, $\sigma$, and $\bm{\theta}$, which can ultimately be expressed by the following equation \citep{bib30}:}
\begin{align}
\begin{matrix}
\hat{\bm{\theta}}=\arg \underset{\bm{\theta}}{\mathop{\max }}\,\left\{ -\frac{n}{2}ln\left( {\bm{\hat{\sigma}}} \right)-\frac{1}{2}ln\left| \mathbf{R} \right| \right\} \\ 
s.t. \begin{cases}
\mathbf{\hat{\bm{\beta}}={{\left( {{F}^{T}}{R}^{-1}F \right)}^{-1}}{{F}^{T}}{R}^{-1}y}\\
{\bm{{{\bm{\hat{\sigma}}}}^{2}}}=\frac{1}{n}\mathbf{{{\left( y-F\hat{\bm{\beta}} \right)}^{T}}{R}^{-1}\left( y-F\hat{\bm{\beta}} \right)}. \\
\end{cases} \\ 
\end{matrix}\label{eq17}
\end{align}

The estimated values of parameters $\hat{\bm{\beta}}$, $\bm{\hat{\sigma}}$, and $\hat{\bm{\theta}}$ can be obtained by solving the optimization problem described above. After that, by combining these estimates into equation (\ref{eq11}), the response values of the unknown points can be calculated.

\subsection{{Motivations and advantages}}\label{subsec2}
For a given stochastic process parameter $\bm{\theta}$, the estimates of the correlation matrix $R$, parameters $\bm{\beta}$, and $\bm{\sigma^2}$ can be easily computed. From equation (\ref{eq4}) of the correlation matrix elements, it can be found that different $\bm{\theta}$ directly affects the prediction results of the Kriging model. {Moreover, since the estimates of parameters $\bm{\beta}$ and $\sigma$ are both influenced by the parameter $\bm{\theta}$, the primary goal of Kriging modelling is to achieve accurate estimation of $\bm{\theta}$.}

\begin{figure*}[h]
\centering
\includegraphics[width=0.95\textwidth]{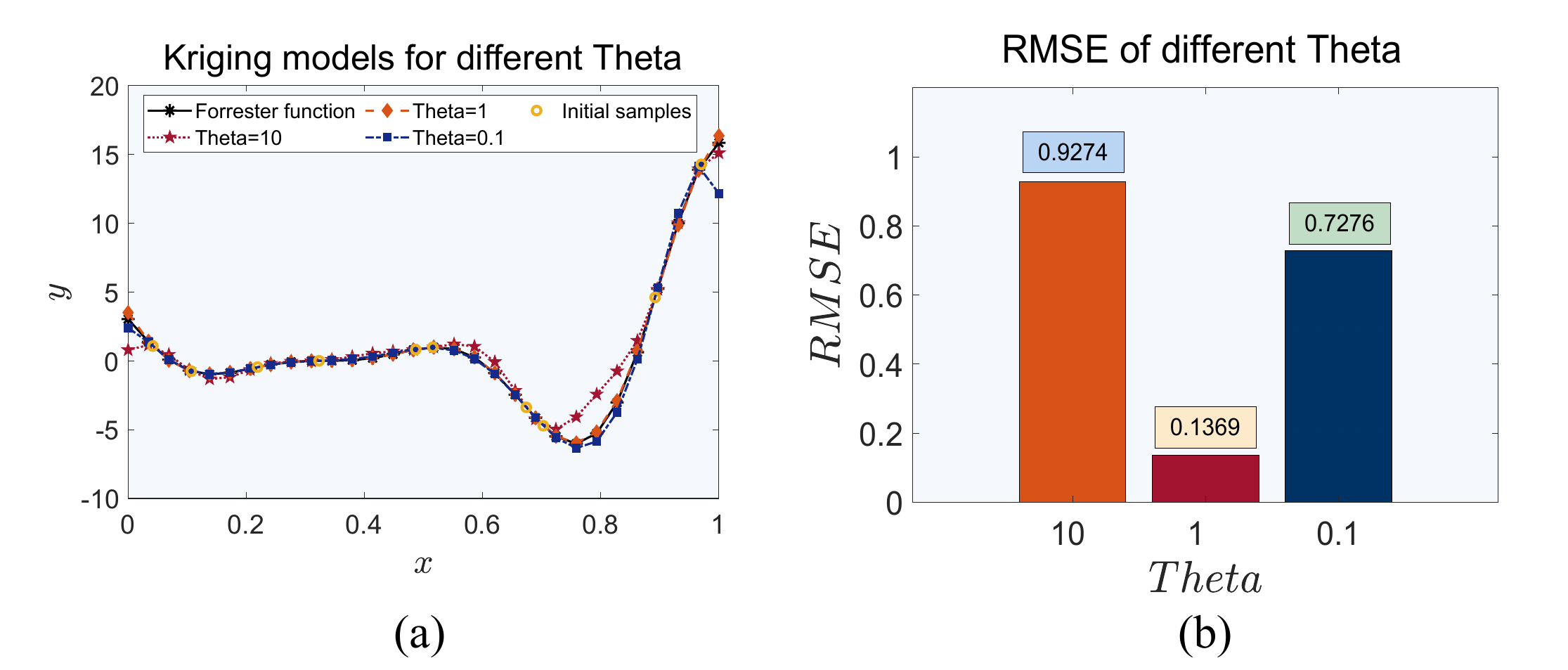}
\caption{Interpolation results of Kriging models with different $\bm{\theta}$}\label{fig2}
\end{figure*}

{To demonstrate the importance of Theta-regularization, we first use the Forrester function to verify the impact of parameter $\bm{\theta}$ on the prediction results of the Kriging model.} The expression of Forrester function \citep{bib42} is shown in equation (\ref{eq18}).
\begin{equation}
f(\mathbf{x})={{(6\mathbf{x}-1)}^{2}}\sin(12\mathbf{x}-4), \mathbf{x}\in [0,1]. \label{eq18}
\end{equation}

{The Latin Hypercube Sampling (LHS) method \citep{bib43} was employed to acquire 10 initial points. LHS is a stratified sampling method used in numerical simulations and experimental design, primarily for generating well-distributed samples within a multi-dimensional parameter space. It divides the sampling space into different layers and then independently draws random samples from these layers.} Simultaneously, the interval [0,1] was discretized, and 30 points were selected for plotting Forrester's original function. Linear and Gaussian functions were chosen as the regression basis function and stochastic process function of the Kriging model, respectively. The results obtained are illustrated in Fig. \ref{fig2}.

{As depicted in Fig. \ref{fig2}, the Kriging model at $\bm{\theta}=1$ exhibits the highest accuracy, with a RMSE (i.e. root mean square error, See equation (\ref{eq37}) in Section \ref{sec4}) of 0.1369}. The predicted model closely aligns with the Forrester function. In contrast, the Kriging model fits at either $\bm{\theta}=10$ or $\bm{\theta}=0.1$ were relatively poor and had a large gap with the actual value of the Forrester function. Accordingly, the Gaussian process hyperparameter $\bm{\theta}$ does have a considerable effect on the accuracy of the Kriging model. Several optimization algorithms have been proposed to determine the optimal value of $\bm{\theta}$. One of the most notable methods is the improved Hooke-Jeeves search algorithm, as implemented in the DACE toolbox \citep{bib35}. This algorithm is widely recognized for its effectiveness in exploring the parameter space by maximizing the likelihood function. However, in some cases, the objective and log-likelihood functions may be flat due to the near-optimal value of the Kriging model. This situation can lead to a significant bias when estimating the parameter $\bm{\theta}$.

In fact, to compute $\bm{\theta}$ for the optimization problem, we need to first obtain the gradient of the likelihood function $L(\bm{\theta})$ at $\bm{\theta}^{(i)}$ using the following analytical form:
\begin{equation}
\frac{\partial L}{\partial {{\bm{\theta}}^{(i)}}}=-\frac{1}{2}\text{tr}( \mathbf{{{R}}^{-1}}\{ {\bm{{\sigma}^{-2}}}\mathbf{(Y-F\bm{\beta}){{(Y-F\bm{\beta})}^{T}}\!-\!{R}}\} 
\mathbf{{{R}}^{-1}}\frac{\partial {\mathbf{R}}}{\partial {{\bm{\theta}}^{(i)}}}),~~for\text{ }i=1,2,\ldots ,D. \label{eq19}
\end{equation} 

Furthermore, use equation (\ref{eq20}) to solve for the element information of the Hessian matrix
\begin{equation}
\begin{split}
{{H}_{ij}}:=&\frac{{{\partial}^{2}}L}{\partial {{\bm{\theta}}^{(i)}}\partial {{\bm{\theta}}^{(j)}}}
=\frac{1}{2}\text{tr}(\mathbf{{{R}}^{-1}}\frac{\partial {\mathbf{R}}}{\partial {{\bm{\theta}}^{(i)}}}\mathbf{{{R}}^{-1}} \{ {\bm{{\sigma}^{-2}}}\mathbf{\left(Y-F\bm{\beta} \right){{\left( Y-F\bm{\beta} \right)}^{T}}-{R}}\}\mathbf{{{R}}^{-1}}\frac{\partial {\mathbf{R}}}{\partial {{\bm{\theta}}^{(j)}}}) \\ 
&\!-\!\frac{1}{2}\text{tr}({\mathbf{R}}^{-1}\{ \bm{{\sigma}^{-2}}\mathbf{\!(Y\!-\!F\bm{\beta}){{(Y\!-\!F\bm{\beta})}^{T}}\!-\!{R}}\}\mathbf{{{R}}^{-1}}\!\frac{{{\partial }^{2}}\mathbf{R}}{\partial {{\bm{\theta}}^{(i)}}\partial {{\bm{\theta}}^{(j)}}}), for\text{ }i,j=1,2,\ldots ,D.
\end{split}  \label{eq20} 
\end{equation}
where $D$ is the dimension, and $\text{tr(}\cdot \text{)}$ is the trace of the matrix.

\begin{figure*}[h]
\centering
\includegraphics[width=0.85\textwidth]{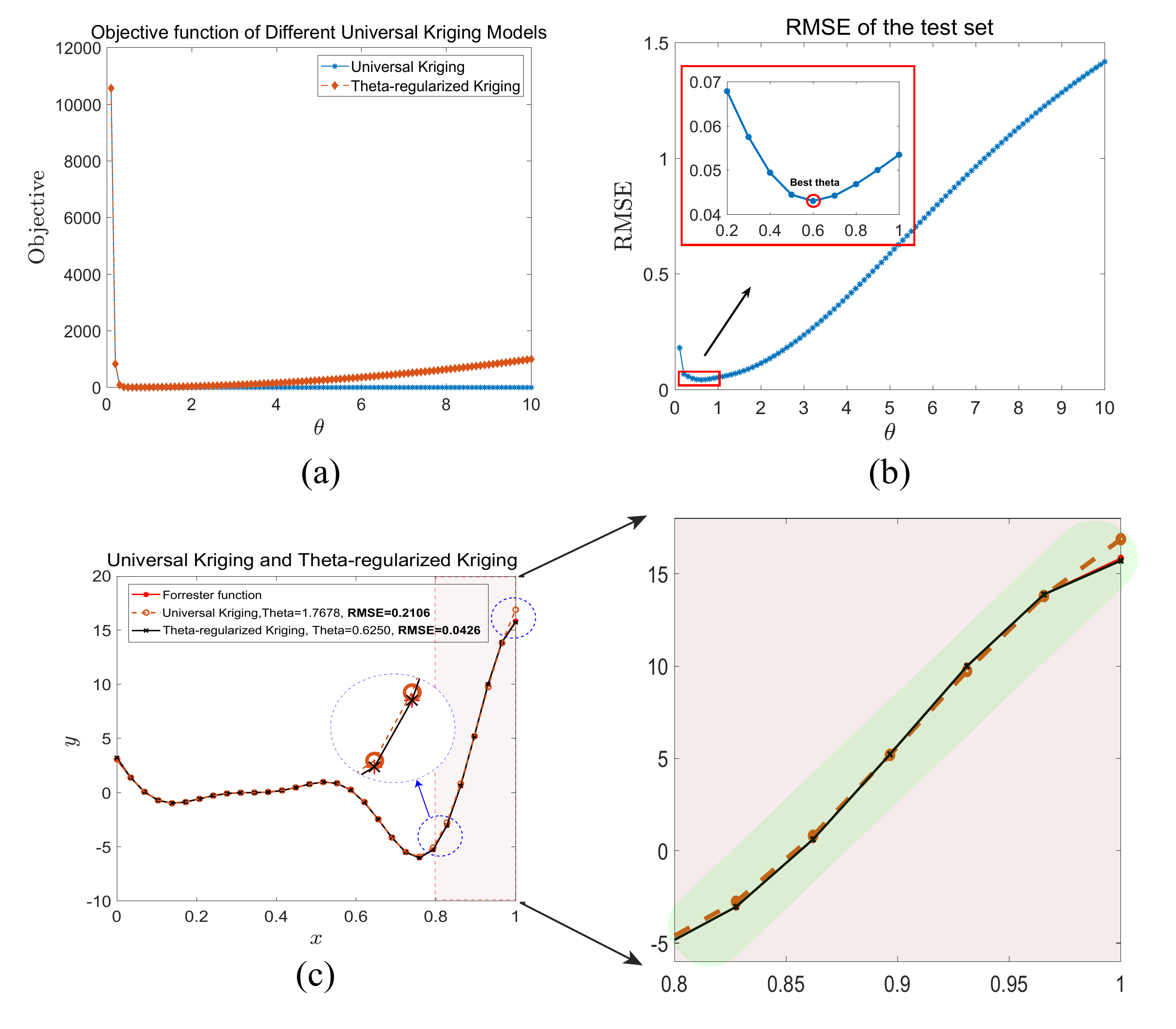}
\caption{Universal Kriging and Theta-regularized Kriging model}\label{fig3}
\end{figure*}

Unfortunately, the complexity of the Kriging model increases significantly as the problem dimension increases, leading to greater computational cost but lower efficiency. {For example, when considering the 8-dimensional Trid function:
\begin{equation}
{f\left( \mathbf{x} \right)=\sum\limits_{i=1}^{8}{{{({\mathbf{x}_{i}}-1)}^{2}}}-\sum\limits_{i=2}^{8}{{\mathbf{x}_{i}}{\mathbf{x}_{i-1}}}.} \label{eqtr}
\end{equation}
{where ${\mathbf{x}_{i}}\in \left[\text{-1, 1} \right],i=\text{1, 2,}...\text{,8}$.}} We calculate its Hessian matrix, and perform a singular value decomposition (SVD) on the matrix to obtain all the singular values, yielding the following results:
\begin{align}
S &=diag(s_1,s_2,...,s_8)\nonumber \\
&=diag(16.6916, 11.1825,..., 0.1323, 0.0053). \nonumber
\end{align}

It can be found that the maximum singular value of the Trid function is four orders of magnitude greater than the minimum singular value. Consequently, the computational efficiency of hyperparameter optimization becomes significantly compromised, leading to suboptimal results. A mature and recognized method for solving this problem is to penalize the loss function or logarithmic likelihood function of the objective model, which is also known as regularization in fields such as mathematics and machine learning. {Hence, enhancing the computational efficiency and accuracy of solving the parameter $\bm{\theta}$ is our initial motivation for implementing regularization.}

Considering the ramifications of the aforementioned problems, we proposed a regularized Kriging model that penalizes the Gaussian stochastic process parameter $\bm{\theta}$. For convenience, we denote it as the Theta-regularized Kriging model. In fact, the TRK model exhibits significant advantages in many aspects. We still use the Forrester function and the same initial conditions to compare the UK model with the TRK model (the penalty term chosen was the Ridge, and the regularization parameter was set to 10). The optimal $\bm{\theta}$ for the UK model was computed by the optimization algorithm of the DACE, and the optimal theta for the TRK model was computed using our improved algorithm. The interpolation results were recorded in Fig. \ref{fig3}, where Fig. \ref{fig3} (a) shows the variation in the objective function of the two Kriging models when taking different values of $\bm{\theta}$. {Figure \ref{fig3} (b) records the trend of RMSE in the Kriging test set with different parameters $\bm{\theta}$, and Figure \ref{fig3} (c) records the approximation effect of the UK and TRK models.} As shown in Fig. \ref{fig3} (a), it can be observed that the objective function of the UK model varies very smoothly, and it is difficult to find the minimum value of the objective function. In contrast, the TRK model more easily finds the optimal solution of the objective function, which is one of the benefits of implementing regularization in the Kriging model.

Another advantage of the TRK model is that penalizing the likelihood function may yield more accurate optimal solutions for the parameters $\bm{\theta}$. For example, in Fig. \ref{fig3}(c), the UK model yields an optimal $\bm{\theta}$ value of 1.7678, while the TRK model ultimately yields an optimized $\bm{\theta}$ value of 0.6250. We know that the goal of the Kriging model is to minimize the MSE loss. Since the Kriging model is an interpolation model, the final judgment of the model's merit needs to use the information from the test set or consider the cross-validation error of the training set. From Fig. \ref{fig3}(b), the RMSE of the Kriging model on the test set is minimized when $\bm{\theta}$ is approximately 0.6. In the executed experiments, the optimal $\bm{\theta}$ obtained using the regularized Kriging model is 0.6250, which is indeed more accurate than the optimal $\bm{\theta}$ obtained by the UK model. Additionally, as evident from Fig. \ref{fig3}(c), the TRK model exhibits better performance than the UK model. The prediction results of the TRK model almost entirely overlap with the true values of the Forrester function, and the RMSE value is only 0.0426 on the test set, whereas the RMSE value of UK is 0.2106.

Overall, the TRK model not only offers significant advantages in finding the minimum of the objective function but also has the opportunity to obtain more accurate solutions when solving for the optimal parameters of the Kriging model. {In fact, the regularization technique can avoid the overfitting phenomenon of the model, as well as enhance the generalization ability of the target model.} Therefore, it is indeed of considerable significance to penalize the hyperparameters $\bm{\theta}$ of the Kriging model.

\section{{The proposed Algorithms}}\label{sec3}
\subsection{{Theory and algorithms of the TRK}}\label{subsec4}
{In Section \ref{sec2}, we elucidated the fundamental principles of the Kriging model and discussed the motivation and advantages of the TRK model. Next, we will focus on the theoretical derivation and practical implementation of the TRK model.}

{Before beginning, we must keep in mind the likelihood function of the Kriging model, which is given by the following formula \cite{bib29}:}
\begin{equation}
ln\left( L \right)=-\frac{n}{2}ln\left( 2\pi \right)-\frac{n}{2}ln\left( {\bm{{\sigma}^{2}}} \right)-\frac{1}{2}ln\left|\mathbf{R}\right|-\frac{\mathbf{{{\left( y-F\bm{\beta} \right)}^{T}}{{{R}}^{-1}}\left( y-F\bm{\beta} \right)}}{2{\bm{{\sigma}^{2}}}}. \label{eq21}
\end{equation} 

In most cases, regularization is implemented by penalizing the model's loss function and adding a nonnegative term. One advantage of such treatment is that it reduces the solution space of the model, thereby preventing model overfitting and potentially leading to more accurate solutions. To some extent, minimizing the loss function and maximizing the likelihood probability are equivalent. Therefore, in this article, we considered penalizing the log-likelihood function of the Kriging model. In the previous section, we explored the effect of the parameter $\bm{\theta}$ on the Kriging model. To account for this, we subtracted a nonnegative term related to the hyperparameter $\bm{\theta}$ from the log-likelihood function of the Kriging model. Consequently, the log-likelihood function of the TRK model can be expressed as:
\begin{equation}
\begin{split}
L_R(\bm{\theta} ) &=ln{{\left( L\left( \bm{\theta} \right) \right)}_{\max }}-\underset{k=1}{\overset{D}{\mathop \sum}}\,p({\bm{\theta}}^{(k)}) \\
& =-\frac{n}{2}ln\left( 2\pi \right)-\frac{n}{2}ln\left( {\bm{{\sigma}^{2}}} \right)-\frac{1}{2}ln\left|\mathbf{R}\right|-\frac{\mathbf{{{\left(y-F\bm{\beta} \right)}^{T}}{{R}}^{-1}\left(y-F\bm{\beta} \right)}}{2{\bm{{\sigma}^{2}}}}-\underset{k=1}{\overset{D}{\mathop \sum}}\,p({\bm{\theta}}^{(k)}), 
\end{split} \label{eq22}
\end{equation} 

Since ${\bm{{\sigma}^{2}}}\!=\!\frac{1}{n}\mathbf{{{(Y\!-\!F\bm{\beta})}^{T}}{{R}^{-1}}(Y\!-\!F\bm{\beta})}$, thus, the above equation can then be further written as
\begin{equation}
L_R(\bm{\theta})=-\frac{n}{2}\left[ ln\left(2\pi \right)+1 \right]-\frac{n}{2}ln\left( {\bm{{\sigma}^{2}}} \right)-\frac{1}{2}ln\left| \mathbf{R} \right|-\underset{k=1}{\overset{D}{\mathop \sum }}\,p(\bm{\theta}^{(k)}), \label{eq23}
\end{equation} 

Similarly, as $-\frac{n}{2}[\ln(2\pi)+1]$ is only related to the sample size $n$, the maximum likelihood estimate of $\bm{\theta}$ can be obtained
\begin{align}
{{{\hat{\bm{\theta}}}}_{MLE}}&=\arg \underset{\bm{\theta}}{\mathop{\max}}\,\left\{ L_R(\bm{\theta} ) \right\}\nonumber \\
&=\arg \underset{\bm{\theta}}{\mathop{\max}}\,\left\{ -\frac{n}{2}ln({\bm{{\sigma}^{2}}})\!-\!\frac{1}{2}ln|\mathbf{R}|\!-\!\underset{k=1}{\overset{D}{\mathop \sum }}\,p({\bm{\theta}}^{(k)})\right\}.  \label{eq24} 
\end{align}

Eq. (\ref{eq24}) can be simplified as
\begin{equation}
{{\hat{\bm{\theta}}}_{MLE}}=\arg \underset{\bm{\theta}}{\mathop{\min }}\,\left\{ \frac{n}{2}ln\left({\bm{{\sigma}^{2}}}{{\left| \mathbf{R} \right|}^{\frac{1}{n}}} \right)+\underset{k=1}{\overset{D}{\mathop \sum }}\,p({\bm{\theta}}^{(k)}) \right\}.  \label{eq25}
\end{equation}

We can find that the above function is monotonic, and $\frac{n}{2}$ is only related to the number of samples. Therefore, we rewrite equation (\ref{eq25}) using exponential operations
\begin{equation}
{{\hat{\bm{\theta}}}_{MLE}}=\arg \underset{\bm{\theta}}{\mathop{\min}}\,\left\{ {\bm{{\sigma}^{2}}}{{|\mathbf{R}|}^{\frac{1}{n}}}\cdot \exp \left\{\frac{2}{n}\underset{k=1}{\overset{D}{\mathop \sum}}\,p({\bm{\theta}}^{(k)}) \right\} \right\}.  \label{eq26}
\end{equation}

It can be observed in Eq. (\ref{eq26}) that $\frac{2}{n}\underset{k=1}{\overset{D}{\mathop \sum }}p({\bm{\theta}}^{(k)})$ tends to become 0 when $n$ is large. However, the Kriging method is a method that uses a small number of samples to replace a high-precision model, and its sample size $n$ is often relatively small. Hence, to avoid the exponential growth of the above function, the penalty coefficient of the TRK model cannot be set too large. This situation is undesirable as it constrains the search space for penalty coefficients, leading to suboptimal parameter solutions. Interestingly, we can eliminate the effect of exponentiality by transformations. It can be noticed that the penalty term $p({\bm{\theta}}^{(k)})=\lambda {{\left| {{\bm{\theta}}^{(k)}} \right|}^{m}}$ is closely related to the penalty parameter $\lambda$, and this parameter is a quantity that we can control. Thus, we can let
\begin{align}
& {\bm{{\sigma}^{2}}}{{|\mathbf{R}|}^{\frac{1}{n}}}\cdot \exp \left\{ \frac{2}{\text{n}}\underset{k=1}{\overset{D}{\mathop \sum}}\,p({\bm{\theta}}^{(k)}) \right\}\!=\!{\bm{{\sigma}^{2}}}{{\left| \mathbf{R} \right|}^{\frac{1}{n}}}\!+\!\tilde{\lambda}\underset{k=1}{\overset{D}{\mathop \sum }}\,{{\left| {{\bm{\theta}}^{(k)}} \right|}^{m}}\nonumber \\ 
& \Leftrightarrow \tilde{\lambda}=\frac{{\bm{{\sigma}^{2}}}{{\left| \mathbf{R} \right|}^{\frac{1}{n}}}\left[ \exp \left\{ \frac{2}{\text{n}}\underset{k=1}{\overset{D}{\mathop \sum }}\,p({\bm{\theta}}^{(k)}) \right\}-1 \right]}{\underset{k=1}{\overset{D}{\mathop \sum }}\,{{\left| {{\bm{\theta}}^{(k)}} \right|}^{m}}}.  \label{eq27}
\end{align}

Therefore, the new penalty term is equal to
\begin{equation}
\underset{k=1}{\overset{D}{\mathop \sum }}\,\tilde{p}({\bm{\theta}}^{(k)})=\tilde{\lambda}\underset{k=1}{\overset{D}{\mathop \sum }}\,{{\left| {{\bm{\theta}}^{(k)}} \right|}^{m}}. \label{eq28}
\end{equation}

According to the above derivation, the optimal coefficient $\bm{\theta}$ of the TRK model is ultimately equivalent to solving the following function:
\begin{equation}
\underset{\bm{\theta}}{\mathop{\min}}\,\left\{ {\bm{{\sigma}^{2}}}{{\left| \mathbf{R} \right|}^{\frac{1}{n}}}+\underset{k=1}{\overset{D}{\mathop \sum}}\,\tilde{p}({\bm{\theta}}^{(k)}) \right\}. \label{eq29}
\end{equation}
where $\underset{k=1}{\overset{D}{\mathop \sum }}\,\tilde{p}({\bm{\theta}}^{(k)})=\tilde{\lambda }\underset{k=1}{\overset{D}{\mathop \sum}}\,{{\left| {{\bm{\theta}}^{(k)}} \right|}^{m}}$, $\tilde{\lambda }$ is the new regularization coefficient, which is not subject to exponential growth but still needs to be given manually or chosen using an algorithm. Essentially, equation (\ref{eq29}) is equivalent to imposing a penalty term on the objective function of the Kriging model.

Consequently, the problem of solving the TRK model can finally be expressed by the following equation:
\begin{equation}
\begin{aligned}
\hat{\bm{\theta}}&=\arg \underset{\bm{\theta}}{\mathop{\min }}\,\left\{ {\bm{{\sigma}^{2}}}{{\left| \mathbf{R} \right|}^{\frac{1}{n}}}+\underset{k=1}{\overset{D}{\mathop \sum}}\,\tilde{p}\left({{\bm{\theta}}^{(k)}} \right) \right\} \\ 
& s.t. \begin{cases}
\mathbf{\hat{\bm{\beta}}={{\left( {{F}^{T}}{R}^{-1}F \right)}^{-1}}{{F}^{T}}{R}^{-1}y}\\
{\bm{{{\bm{\hat{\sigma}}}}^{2}}}=\frac{1}{n}\mathbf{{{(y-F\hat{\bm{\beta}})}^{T}}{R}^{-1}(y-F\hat{\bm{\beta}})}. \\
\end{cases}
\end{aligned}  \label{eq30} 
\end{equation}

\begin{algorithm*}[h]
\caption{Iterative Process of the TRK Model.}\label{algo1}
\begin{algorithmic}[1]
\STATE \textbf{Input:} Design points $(\mathbf{x}_i,\mathbf{y}_i)$, initial value ${\bm{\theta}_0}$, maximum of iterations ${i_{max}}$, objective function difference threshold $\epsilon$, penalty function $\tilde{p}({\bm{\theta}})$ and its coefficients.
\STATE \textbf{Output:} $\hat{\bm{\theta}}$, $\hat{\bm{\beta}}$, ${\bm{{{\bm{\hat{\sigma}}}}^{2}}}$, and $\hat{\mathbf{y}}(\mathbf{x})$.
\STATE Let $i=0$, and set the initial value of parameter ${{\bm{\theta}}_{0}}$ of the stochastic process.
\STATE \textbf{while} {$i < i_{max}$}
\STATE \quad Calculate the correlation matrix $\mathbf{R}({{\bm{\theta}}_{i}})$.
\STATE \quad Solve Eq.(\ref{eq30}) to obtain the coefficients $\bm{\beta}$ and ${\bm{{\sigma}^{2}}}$.
\STATE \quad Solve the equation for ${{\hat{\bm{\theta}}}_{MLE}}=\arg \underset{\bm{\theta}}{\mathop{\min }}\,\left\{ {\bm{{\sigma}^{2}}}{{\left| \mathbf{R} \right|}^{\frac{1}{n}}}+\underset{k=1}{\overset{D}{\mathop \sum }}\,\tilde{p}({\bm{\theta}}^{(k)}) \right\}$.
\STATE \quad Update the relevant parameter ${{\bm{\theta}}_{i+1}}$.
{\STATE   \quad \textbf{if} $\mathcal{L}_{i+1}-\mathcal{L}_{i}<\epsilon$
\STATE    \quad \quad break.
\STATE  \quad \textbf{else} 
\STATE   \quad \quad Let $i=i+1$, and continue the iteration.}
\STATE \textbf{end while}.
\STATE Save the final iteration values of parameters $\hat{\bm{\theta}}$, $\hat{\bm{\beta}}$, and ${\bm{{{\bm{\hat{\sigma}}}}^{2}}}$.
\STATE Substitute these parameters into Eq.(\ref{eq11}) to predict the response values of unknown points $\hat{\mathbf{y}}(\mathbf{x})$.
\end{algorithmic}
\end{algorithm*}

There are many types of regularization, and in this research we considered three common regularization penalty terms, namely, the Lasso penalty (also known as ${{\ell}_{1}}$ regularization), Ridge penalty (also known as ${{\ell}_{2}}$ regularization), and Elastic-net penalty. The relevant definitions of these penalties are as follows:
\begin{equation}
{{\ell}_{1}}:{{p}_{\lambda}}({\bm{\theta}}^{(k)})=\lambda \left| {{\bm{\theta}}^{(k)}} \right|, \label{eq31}
\end{equation}
\begin{equation}
{{\ell}_{2}}:{{p}_{\mu }}({\bm{\theta}}^{(k)})=\mu \left| {{\bm{\theta}}^{(k)}} \right|^{2}, \label{eq32}
\end{equation}
\begin{equation}
\text{Elastic-net}:{{p}_{\gamma}}({\bm{\theta}}^{(k)})=\gamma \cdot[\alpha |{{\bm{\theta}}^{(k)}}|+(1-\alpha) | {{\bm{\theta}}^{(k)}}|^{2}], \label{eq33}
\end{equation}
where $\lambda$ and $\mu$ are the penalty coefficients of ${{\ell }_{1}}$ and ${{\ell }_{2}}$ regularization respectively; $\gamma$ and $\alpha$ is the penalty coefficient and weighting coefficient of Elastic-net regularization. The Elastic-net penalty is essentially a weighted combination of the Lasso and Ridge penalties, which regulates the ratio of ${{\ell }_{1}}$ and ${{\ell }_{2}}$ regularization through the weighting coefficient $\alpha$. When $\alpha=0$, the Elastic-net penalty is the Ridge penalty; when $\alpha=1$, it is the Lasso penalty.

\begin{figure*}[h]
\centering
\includegraphics[width=1\textwidth]{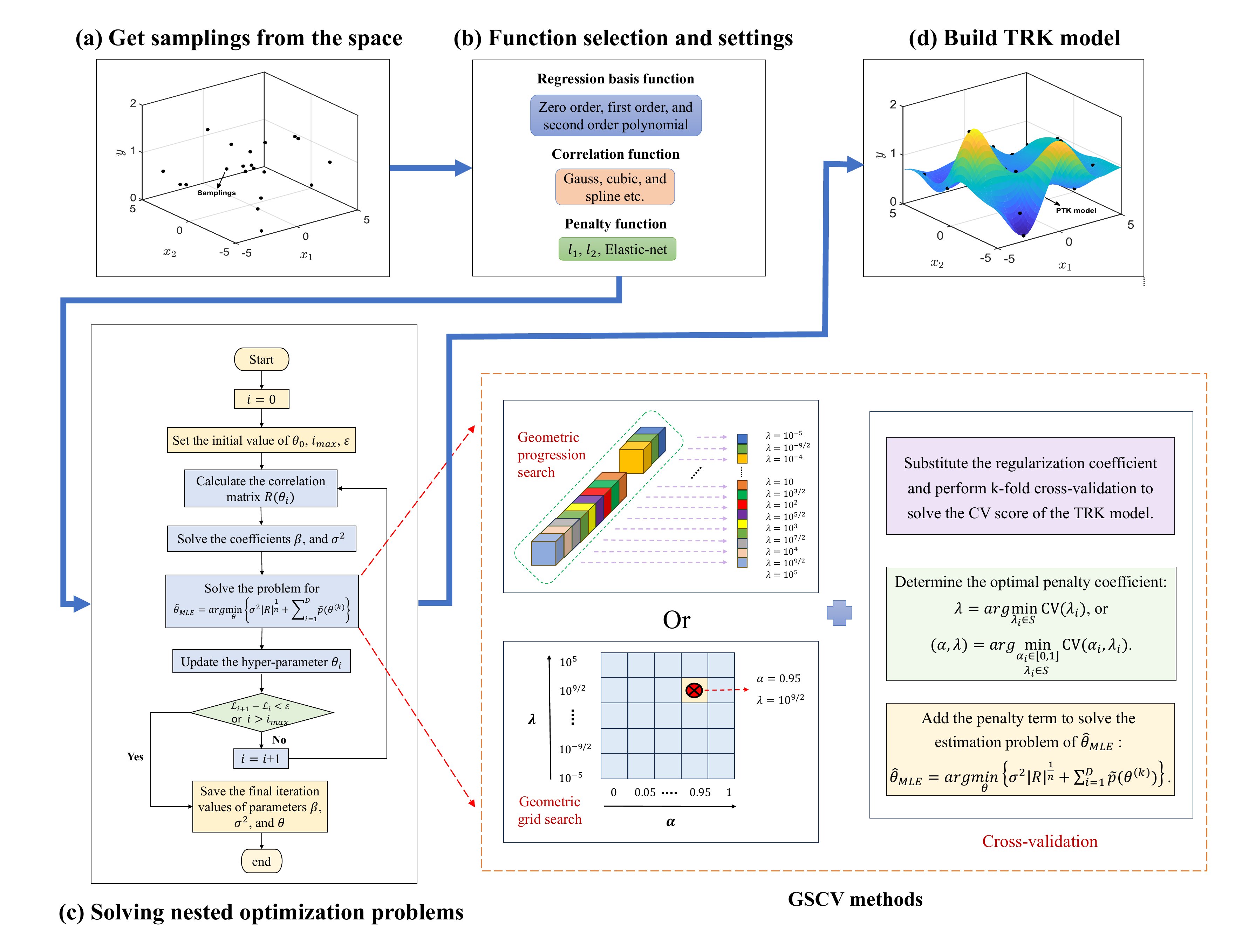}
\caption{Illustration of the proposed TRK model. Firstly, samplings are obtained from the sample space. Subsequently, regression basis functions, correlation functions, and penalty functions are defined. Following this, a nested optimization problem needs to be solved, ultimately building the TRK model.}\label{figmain}
\end{figure*}

The TRK model is still essentially a nested optimization problem for solving the parameters $\bm{\beta}$, ${\bm{{\sigma}^{2}}}$ and $\bm{\theta}$. Therefore, an iterative algorithm is required to estimate the parameters of this model. The construction process of the TRK model is outlined in Figure \ref{figmain}, and the iterative algorithm for the TRK model is summarized in Algorithm \ref{algo1}.

\subsection{{Parameter sensitivity analysis of the TRK model}}\label{subsubsec2}
\begin{figure*}[h]
\centering
\includegraphics[width=0.95\textwidth]{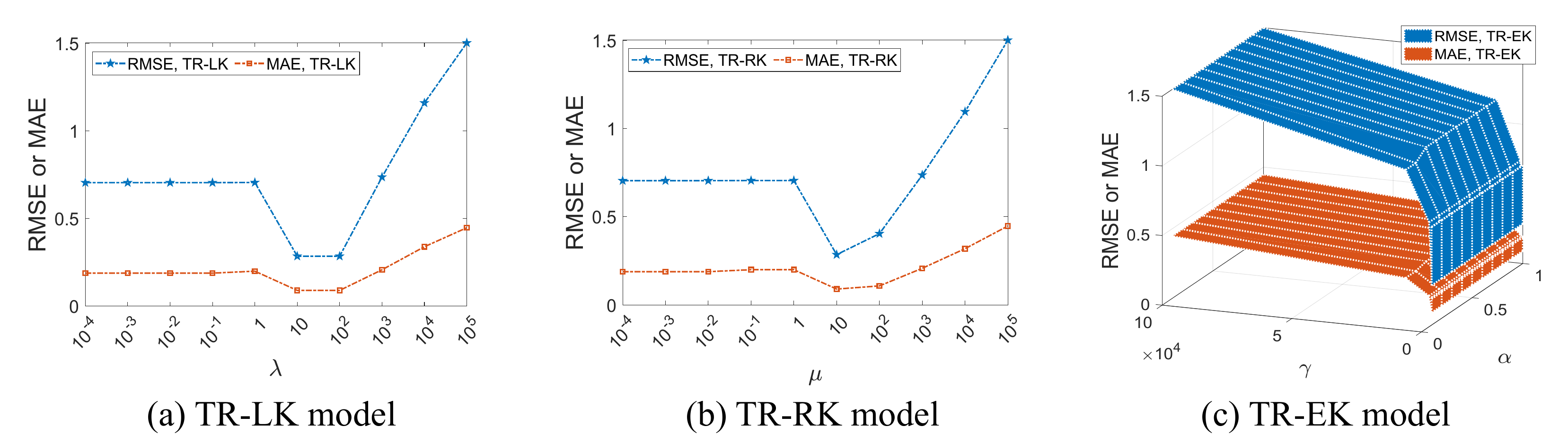}
\caption{Sensitivity of TRK models to different parameters.} \label{figsens}
\end{figure*}

{Due to the addition of regularization penalties in the TRK model, it is necessary to evaluate the sensitivity of the TRK model to different penalty parameters. When considering three different penalty TRK models, the Theta-regularized Lasso Kriging (TR-LK) and Theta-regularized Ridge Kriging (TR-RK) models each have only one regularization parameter. Therefore, various penalty parameters are selected, and the Forrester function from Section \ref{subsec2} is used to analyze the effect of $\lambda$ and $\mu$ on the accuracy of the TR-LK and TR-RK models. RMSE and MAE (i.e. mean absolute error, See equation (\ref{eq38}) in Section \ref{sec4}) were used to reflect the accuracy of the model. The analysis results are shown in Figure \ref{figsens}(a) and \ref{figsens}(b). The penalty parameters of different magnitudes have different impacts on the accuracy of TR-LK and TR-RK models. It can be observed that for this test case, the TR-LK model and TR-RK model seem to have high accuracy when the penalty parameter is set to 10. When the penalty parameter is set to other magnitudes, the accuracy of the model will vary by different magnitudes. Therefore, when adjusting the regularization parameters in practice, it is recommended to consider a geometric sequence such as ${10^{-5}, 10^{-4}, ..., 10^5}$.}

{In the construction of the Theta-regularized Elastic-net Kriging (TR-EK) model, as indicated by Equation (\ref{eq33}), the new parameters introduced are $\alpha$ and $\gamma$. Due to the involvement of two parameters, a grid search method is considered for parameter selection, with the results shown in Figure \ref{figsens}(c). It is evident that the impact of parameters $\alpha$ and $\gamma$ on model accuracy varies significantly. In this case, when the $\gamma$ value is fixed, changes in the $\alpha$ parameter seem to have a minor effect on model accuracy, whereas changes in the $\gamma$ parameter, when $\alpha$ is fixed, have a substantial impact on model accuracy. Therefore, for the TR-EK model, the regularization parameter $\gamma$ is more sensitive than the weight parameter $\alpha$. Additionally, the effect of different magnitudes of $\gamma$ on model accuracy varies greatly. Thus, when adjusting the parameters of the TR-EK model, priority should be given to exploring different magnitudes of the $\gamma$ parameter, followed by selecting a suitable range for the weight parameter $\alpha$.}

\subsection{{GSCV algorithm for solving optimal regularization parameters}}
{According to the results of parameter sensitivity analysis, in practical applications, we can typically manually select the regularization coefficient of the TRK model in the form of a geometric progression.} However, considering that manually selected regularization coefficients are subjective and require constant testing experiments, we propose a method that combines geometric progression search and cross-validation to obtain the optimal regularization coefficients, named the GSCV method. This method can more accurately and effectively identify the optimal penalty coefficient when the parameters are set reasonably. Moreover, it is noteworthy that, in the case of Elastic-net regularization, the search method is geometric grid search. The difference is that Elastic-net has an additional weight parameter $\alpha$ that needs to be determined.

The principle of cross-validation (CV) is defined as follows:

\textit{Step} 1. Split the data $D$ into V subsets ${{D}_{1}},{{D}_{2}},\ldots ,{{D}_{V}}$, where $D\!=\!\left\{(\mathbf{x}_{i},\mathbf{y}_{i})\!:i\!=\!1,\ldots ,n \right\}$.

\textit{Step} 2. For $k=1,\ldots ,V$, let ${{D}^{\left( k \right)}}=D-{{D}_{k}}$. Then, build the model using dataset ${{D}^{\left( k \right)}}$, and use subset ${{D}^{\left( k \right)}}$ for prediction to obtain the predicted value ${{\hat{\mathbf{y}}}^{\left( k \right)}}(\mathbf{x})$.

\textit{Step} 3. Calculate the MSE value of the CV.
\begin{equation}
\text{CV}\left( \text{ }\!\!\lambda\!\!\text{ } \right)=\frac{\mathop{\sum }_{v}\mathop{\sum }_{\left({\mathbf{x}_{i}},{\mathbf{y}_{i}} \right)\in {{D}_{v}}}\frac{{{\left( {\mathbf{y}_{i}}-{{{\hat{\mathbf{y}}}}^{\left( v \right)}}\left( \mathbf{x}_{i} \right) \right)}^{2}}}{size({{D}_{k}})}}{V}, \label{eq34}
\end{equation}
where $size({{D}_{k}})$ is the number of elements in subset ${{D}_{k}}$.

\textit{Step} 4: For a given set of $S=\left\{ {{\lambda }_{1}},{{\lambda }_{2}},\ldots ,{{\lambda }_{k}} \right\}$, the best regularization parameter $\lambda $ corresponding to the smallest CV score.
\begin{equation}
\lambda =\arg \underset{{{\lambda }_{i}}\in S}{\mathop{\min}}\,\text{CV}\left( {{\lambda }_{i}} \right). \label{eq35}
\end{equation}

We know that the regularization coefficient is generally greater than 0, and coefficients of the same magnitude do not significantly affect model penalization. Therefore, using a geometric progression to find the optimal regularization coefficient is a good choice. Let the first term of the geometric progression be denoted as $a_0$, the common ratio as $q$, and the number of iterations as $n$. Thus, the proposed GSCV method can determine the optimal penalty coefficient by providing $a_0$, $q$, $n$, and $k$ (the fold of cross-validation). Theoretically, the smaller the common ratio $q$ or the larger the number of iteration terms $n$, the better the result of the selection of regularization coefficients. Overall, the algorithm of the proposed GSCV method is shown in Algorithm \ref{algo2}.

\begin{algorithm*}
\caption{GSCV method for selecting regularization coefficients.}\label{algo2}
\begin{algorithmic}[1]
\STATE \textbf{Input:} Design points $\mathbf{(x_i,y_i)}$, GSCV method parameters: $k$, $a_0$, $q$, $n$ (and $\alpha, h$ for Elastic-net regularization).
\STATE \textbf{Output:} The optimal regularization coefficient $\lambda$.
\STATE Set the values of the cross-validation fold number $k$, the $a_0$, $q$, $n$ of the geometric progression, and the Elastic-net coefficients $\alpha,h$ (if needed).
\STATE let $i = 1$, use geometric progression (or geometric grid) search and cross-validation method to execute the iterative process.
\STATE \textbf{while} {$i < n+1$}
\STATE Calculate the regularization coefficient $\lambda_i$ and the weight coefficient $\alpha_i$ (Elastic-net regularization), where $\lambda_i$ = ${{a}_{0}}{{q}^{i-1}}$, $\alpha_i=(i-1)h$.
\STATE Substitute the regularization coefficient and perform $k$-fold cross-validation to solve the CV score of the regularized Kriging model.
\STATE \textbf{end while}.
\STATE Determine the optimal regularization coefficient $\lambda =\arg \underset{{{\lambda}_{i}}\in S}{\mathop{\min}}\,\text{CV}\left( {{\lambda}_{i}} \right)$.
\end{algorithmic}
\end{algorithm*}

In this study, we developed a MATLAB software package named OptRP based on the principle of the GSCV method to optimize the selection of regularization parameters. The OptRP package can be found on the website \url{https://github.com/xuelin-xie/Regularized-Kriging}. When using the GSCV method to obtain optimal penalty coefficients, we set the number of iteration terms and the common ratio to $n = 20$ and $q = 10^{1/2}$, respectively. The first term of the iteration is $a_0 = 10^{-5}$, and the folds for cross-validation are $k = 5$. This setting enables a search for relatively good penalty coefficients within a reasonable range and is also the default value in our OptRP package. Essentially, this setting is equivalent to searching for optimal iteration coefficients on the interval $[10^{-5}, 10^5]$ with $q = 10^{1/2}$ as the common ratio. Users can also define the first term, the number of iteration terms, and the common ratio of the OptRP package themselves, depending on their specific needs.

\section{{Numerical examples}}\label{sec4}
{In this section, we validate the superior accuracy and stability of the TRK model using 10 numerical test functions and 2 practical engineering problems. In addition to the RMSE and MAE used in the previous sections, the coefficient of determination (${\text{R}^2}$) was also chosen as a metric for model evaluation. Among these, ${\text{R}^2}$ is used to reflect overall performance, RMSE indicates global accuracy, and MAE represents the local accuracy of the model. The expressions for these evaluation metrics are as follows:}
\begin{equation}
\text{R}^{2}=1-\frac{\mathop{\sum }_{i=1}^{n}{{\left( {\mathbf{y}_{i}}-{{{\hat{\mathbf{y}}}}_{i}} \right)}^{2}}}{\mathop{\sum }_{i=1}^{n}{{\left( {\mathbf{y}_{i}}-\bar{\mathbf{y}} \right)}^{2}}},   \label{eq36}
\end{equation}
\begin{equation}
\text{RMSE}=\sqrt{\frac{\mathop{\sum }_{i=1}^{n}{{\left( {\mathbf{y}_{i}}-{{{\hat{\mathbf{y}}}}_{i}} \right)}^{2}}}{n}}, \label{eq37}
\end{equation}
\begin{equation}
\text{MAE}=\frac{\sum\nolimits_{i=1}^{n}{\left| {\mathbf{y}_{i}}-{{{\hat{\mathbf{y}}}}_{i}} \right|}}{n}, \label{eq38}
\end{equation}
where ${\mathbf{y}_{i}}$ is the true response value, ${\hat{\mathbf{y}}_i}$ is the predicted value, and ${\bar{\mathbf{y}}}$ is the mean of ${\mathbf{y}_{i}}$.

\subsection{Test functions}\label{subsec5}
\subsubsection{Experimental setup and numerical functions}\label{subsubsec3}
{Nine numerical benchmark functions were selected to test the accuracy and stability performance of TRK, UK, and other PBK \citep{bib29,bib30} and MPBK \citep{bibzhao} models. {The abbreviations for different regularized Kriging models are provided in Table \ref{tab1}. The optimal penalty coefficients of these models were obtained by using our proposed GSCV method. The ranges for the regularization parameters $\lambda$, $\mu$, and $\gamma$ are $[10^{-5}, 10^{5}]$ with a ratio of $q = 10^{1/2}$. Additionally, for the Elastic-net penalty, the weight coefficient $\alpha$ ranges from 0 to 1, with a step size of $h = 0.05$.} The dimensions of the input variables for the nine selected benchmark functions range from 2 to 24 dimensions, and their function expressions are shown in equations (\ref{eq39}) - (\ref{eq48}).}
\\\\
1. Corner Peak function
\begin{equation}
f(\mathbf{x})={{(1+5({\mathbf{x}_{1}}+{\mathbf{x}_{2}}))}^{-3}}. \label{eq39}
\end{equation}
{where ${\mathbf{x}_{i}}\in \left[ \text{0, 1}\right],i=\text{1, 2}$.} \\
2. Langermann function
\begin{equation}
f(\mathbf{x})\!=\!\sum\limits_{i=1}^{5}{{{c}_{i}}\exp(\!-\!\frac{1}{\pi}\sum\limits_{j=1}^{2}{{{(\mathbf{x}_{j}\!-\!{A}_{ij})}^{2}}})}\cos(\pi\sum\limits_{j=1}^{2}{{{(\mathbf{x}_{j}\!-\!{A}_{ij})}^{2}}}),  \label{eq41}
\end{equation}
{where $c=(1,2,5,2,3),A={{\left( \begin{matrix}
   3 & 5 & 2 & 1 & 7  \\
   5 & 2 & 1 & 4 & 9  \\
\end{matrix} \right)}^{T}},{\mathbf{x}_{j}}\in \left[ \text{0, 1} \right],j=\text{1, 2}$.} \\
3. {Rastrigin function
\begin{equation}
f\left(\mathbf{x} \right)=20+\underset{i=1}{\overset{2}{\mathop \sum }}\,\left[\mathbf{x}_{i}^{2}-10cos\left( 2\pi {\mathbf{x}_{i}} \right) \right], \label{eq40}
\end{equation}
{where ${\mathbf{x}_{i}}\in \left[0,1.8 \right],i=1,2$.} \vspace{2pt}}\\
4. Morokoff and Caflisch function
\begin{equation}
f\left( \mathbf{x} \right)=\frac{9}{4}\prod\limits_{i=1}^{2}{{{({\mathbf{x}_{i}})}^{1/2}}}. \label{eq42}
\end{equation}
{where ${\mathbf{x}_{i}}\in \left[ \text{0, 1} \right],i=\text{1, 2}$.} \\
5. Sphere function
\begin{equation}
f\left(\mathbf{x} \right)=\sum\limits_{i=1}^{4}{\mathbf{x}_{i}^{2}}. \label{eq43}
\end{equation}
{where ${\mathbf{x}_{i}}\in \left[ \text{-5}\text{.12, 5}\text{.12} \right],i=\text{1, 2,}...\text{,4}$.} \\
6. RHE function
\begin{equation}
f\left( \mathbf{x} \right)=\mathop{\sum }_{i=\text{1}}^{\text{6}}\mathop{\sum }_{j=\text{1}}^{i}\mathbf{x}_{j}^{2}. \label{eq44}
\end{equation}
{where ${\mathbf{x}_{j}}\in \left[ \text{-1, 1} \right],i=\text{1, 2,}...\text{,6,} j=1,2,...,i$.} \\
7. Trid function
\begin{equation}
f\left( \mathbf{x} \right)=\sum\limits_{i=1}^{8}{{{({\mathbf{x}_{i}}-1)}^{2}}}-\sum\limits_{i=2}^{8}{{\mathbf{x}_{i}}{\mathbf{x}_{i-1}}}. \label{eq45}
\end{equation}
{where ${\mathbf{x}_{i}}\in \left[ \text{-1, 1} \right],i=\text{1, 2,}...\text{,8}$.} \\
8. Schwefel function
\begin{equation}
f\left( \mathbf{x} \right)=418.9829\times 12-\sum\limits_{i=1}^{12}{{\mathbf{x}_{i}}\sin (\sqrt{|{\mathbf{x}_{i}}|})}. \label{eq46}
\end{equation}
{where ${\mathbf{x}_{i}}\in \left[ \text{-1, 1} \right],i=\text{1, 2,}...\text{,12}$.} \\
9. Stybtang function
\begin{equation}
f\left( \mathbf{x} \right)=\frac{1}{2}\sum\limits_{i=1}^{24}{(\mathbf{x}_{i}^{4}-16\mathbf{x}_{i}^{2}+5{\mathbf{x}_{i}})}. \label{eq48}
\end{equation}
where ${\mathbf{x}_{i}}\in \left[ \text{0, }{1}/{2}\right], i=\text{1, 2,}\ldots ,\text{24}$. \\

\begin{figure*}[h]
\centering
\includegraphics[width=0.9\textwidth]{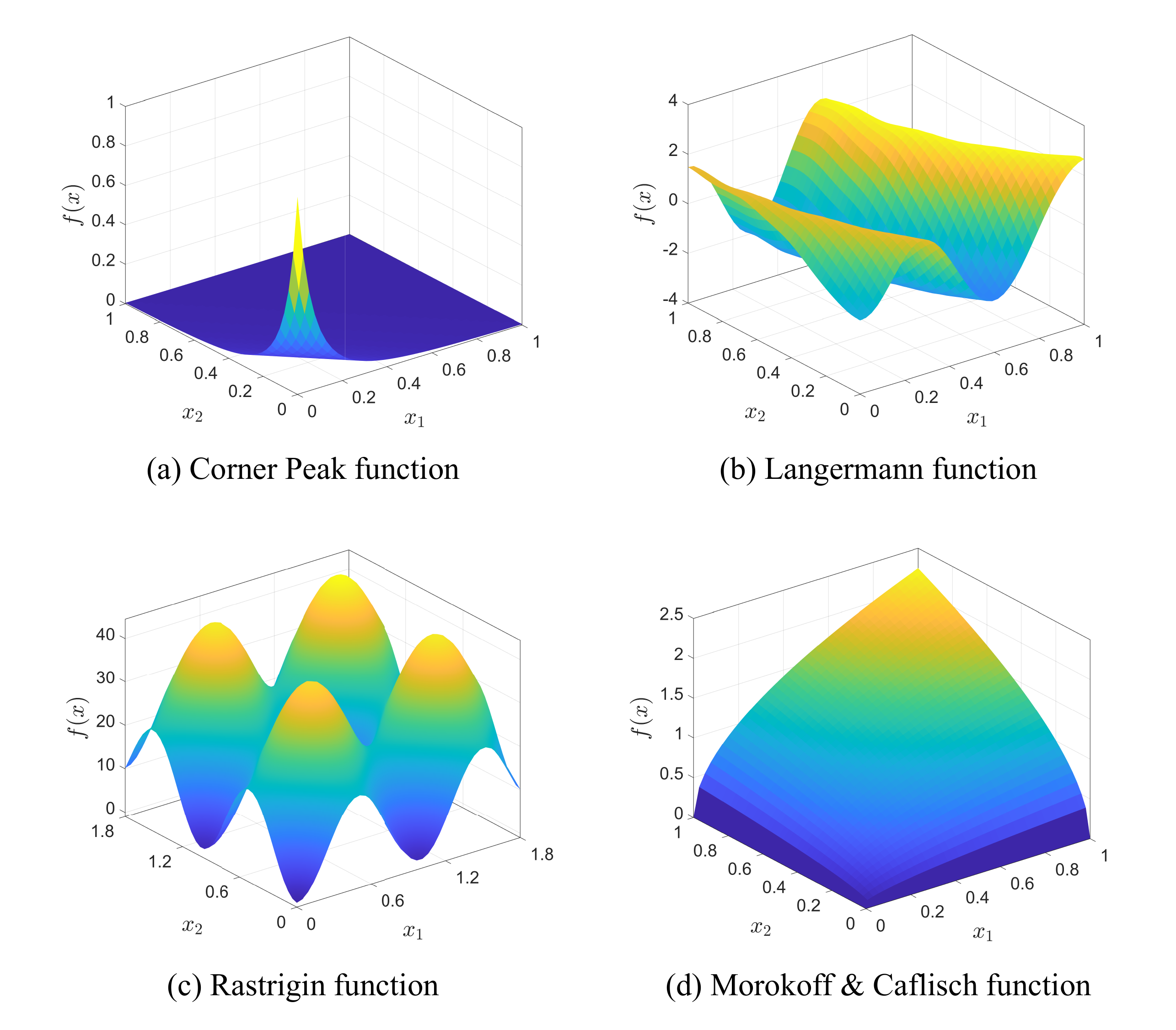}
\caption{Schematic diagram of the two-dimensional test functions}\label{fig4}
\end{figure*}

{In these test functions, the surfaces of the two-dimensional test functions are depicted in Figure \ref{fig4}. The Corner Peak function \citep{bib44} has a flat surface but rises sharply to a peak at one corner. This function allows for rapid analytical integration with high precision. The Langermann function \citep{bib45} is multimodal and exhibits fewer local minima but greater nonlinearity within the interval [0,1] compared to the Corner Peak function. {The Rastrigin function is a non-convex function \citep{bib46}, which is characterized by its high oscillation, multiple peaks, and numerous local minima.} The final two-dimensional function is the Morokoff and Caflisch function \citep{bib47}. When used as an integrand, the integral value of the Morokoff and Caflisch function is exactly 1, with its variance and variation easily obtained since it is formed by the product of one-dimensional functions.}

{In addition to the two-dimensional functions, we also considered higher-dimensional cases. The Sphere function and the Rotated Hyper-Ellipsoid (RHE) function are both continuous, convex, and unimodal \citep{bib48}. Besides the global minimum, the Sphere function has d local minima (where d is the dimension). The RHE function, also known as the sum of squares function, is an extension of the axis-parallel hyper-ellipsoid function. The Trid function has no local minima other than the global minimum. In contrast, the Schwefel function is complex and has many local minima. Lastly, we considered a medium-to-high dimensional numerical function: the 24-dimensional Stybtang function \citep{bib50}. This benchmark function is a commonly used test function in intelligent optimization algorithms and surrogate modeling. Due to its nonlinear and multimodal properties, it was also considered in this study.}

For each test function, we used LHS to collect 60 and 90 samples for training the model, and an additional 5000 samples for testing. To ensure fairness, the sampling and validation points were controlled to be the same. In our experiment, the initial value of the parameter $\bm{\theta}$ was set to 10, and the search interval was $[10^{-2}, 10^2]$. The regression function of all Kriging models was set to a linear function, while the correlation function was the Gaussian function.

\begin{table*}[h]
\caption{{Abbreviations and parameters of different regularized Kriging models}} \label{tab1}
\resizebox{0.7\columnwidth}{!}{
\begin{tabular*}{0.85\linewidth}{lll}
\toprule
Abbreviations ~~~~& Regularized models ~~~~& Parameters\\
\midrule
TR-LK ~~~~& Theta-regularized Lasso Kriging ~~~~& $\lambda_{TRLK}$ \\
TR-RK ~~~~& Theta-regularized Ridge Kriging ~~~~& $\mu_{TRRK}$\\
TR-EK ~~~~& Theta-regularized Elastic-net Kriging ~~~~& $\gamma_{TREK}, \alpha_{TREK}$ \\
PB-LK ~~~~& Penalized blind Lasso Kriging ~~~~& $\lambda_{PBK}$ \\
PB-RK ~~~~& Penalized blind Ridge Kriging ~~~~& $\mu_{PBLK}$ \\
PB-EK ~~~~& Penalized blind Elastic-net Kriging ~~~~& $\gamma_{PBEK}, \alpha_{PBEK}$ \\
MPBK  ~~~~& Modified Penalized blind Kriging ~~~~& $\gamma_{MPBK}, \alpha_{MPBK}$ \\
\hline
\end{tabular*} 
}
\end{table*}

\subsubsection{{Numerical results for nine benchmark functions}}\label{subsubsec4}
To avoid significant accidental errors caused by a single experiment, after setting the relevant parameters of the universal and regularized Kriging models, we repeated all experiments ten times and statistically analyzed the mean and standard deviation. The statistical results are recorded in Tables \ref{tab2}-\ref{tab7}, where Tables \ref{tab2}-\ref{tab4} present the results of the UK and regularized Kriging models established at an initial sampling point of 60, and Tables \ref{tab5}-\ref{tab7} present the results of the universal and regularized Kriging models established at an initial sampling point of 90. The best result is displayed in \textbf{bold}, and the suboptimal value is \underline{underlined}.

From Tables \ref{tab2}-\ref{tab7}, it can be noted that for different sampling points and different dimensions of the test function, the proposed TRK model is ahead of the other Kriging models in terms of accuracy and stability. The number of initial sampling points has a certain impact on the performance of the Kriging model. In the cases of fewer sampling points, the TRK models seem to improve accuracy more. Of course, this is not absolute. For instance, for the Trid function, the predictive performance improvement of the TRK models is more significant when the initial sample point is $n=90$ compared to $n=60$. {From Tables \ref{tab2} and \ref{tab5}, we can observe that for different test functions, the $\text{R}^2$ of different Kriging models are mostly close to 1, indicating that the models exhibit good fitting performance when approximating these test functions. Of course, the predictive performance of the models fundamentally depends on the difficulty of the problem. When dealing with low-dimensional or relatively simpler problems, the models can achieve higher accuracy. However, as the problem becomes high-dimensional or more complex, such as with the RHE and Stybtang functions, achieving high accuracy with these models becomes challenging.}

\begin{table*}[h]
\caption{Test results of ${\text{R}^2}$ for 60 training samples (Mean/Std)}
\label{tab2}
\resizebox{1\columnwidth}{!}{
\begin{tabular*}{1.49\linewidth}{lllllllll}
\toprule
Functions & UK & TR-LK & TR-RK & TR-EK & PB-LK & PB-RK & PB-EK & MPBK \\
\midrule
Corner Peak & 0.8621/0.0687 & 0.9385/0.0343 & 0.9284/0.0398  & 0.9205/0.0413 & \textbf{0.94525}/0.0343 & 0.9345/0.0403 & \underline{0.94517}/0.0317 & 0.9250/0.0496\\
Langermann & 0.99529/3.17E-03 & \underline{0.99530}/3.16E-03 & \textbf{0.9954}/3.14E-03 & 0.9915/1.11E-02 & 0.99529/3.17E-03 & 0.8368/4.61E-01 & 0.8493/4.64E-01 & 0.9731/4.91E-02 \\
Rastrigin & 0.9953/0.0032	& \underline{0.9960}/0.0033 	& \textbf{0.9961}/0.0031		& 0.9956/0.0031	& 0.9881/0.0238 	& 0.9093/0.1239	& 0.8515/0.1417  & 0.8728/0.1986\\
Morcaf95a & 0.9958/3.75E-03 & \textbf{0.9990}/8.11E-04 & 0.9988/8.36E-04 & 0.9984/2.03E-03 & 0.99897/8.09E-04 & \underline{0.99898}/8.11E-04	 & \underline{0.99898}/8.11E-04 & 0.9974/2.68E-03\\
Sphere & 0.9897/0.0108 & 0.9978/0.0012 & \textbf{0.9986}/0.0009 & \underline{0.9979}/0.0011 & 0.9850/0.0235 & 0.9867/0.0150 & 0.9831/0.0182 & 0.9958/0.0026\\
RHE & 0.6737/0.1941 & 0.7205/0.0776 & \underline{0.7234}/0.1151 & \textbf{0.7423}/0.0714 & 0.6794/0.2030& 0.6678/0.1509 & 0.6664/0.2002 & 0.5436/0.2724\\
Trid & 0.8802/0.0117 & 0.8827/0.0132  & \textbf{0.8873}/0.0273 & \underline{0.8870}/0.0288 & 0.8725/0.0149 & 0.8703/0.0140 & 0.8685/0.0113 & 0.8741/0.0226\\
Schwefel & 0.9835/1.03E-03 & \underline{0.98379}/1.02E-03 & 0.9837/1.06E-03 & \textbf{0.98383}/9.94E-04 & 0.9835/1.03E-03	& 0.9835/1.03E-03	 & 0.9835/1.03E-03 & 0.9837/1.92E-03\\
Stybtang & 0.4547/0.0945	& \underline{0.4559}/0.0936 & \textbf{0.4567}/0.0927 & 0.4558/0.0938 & 0.4547/0.0945 & 0.4547/0.0945  & 0.4547/0.0945 & 0.4404/0.0867\\
\hline
\end{tabular*}
}
\end{table*}

\begin{table*}[h]
\caption{Test results of RMSE for 60 training samples (Mean/Std)}\label{tab3}
\resizebox{1\columnwidth}{!}{
\begin{tabular*}{1.43\linewidth}{lllllllll}
\toprule
Functions & UK & TR-LK & TR-RK & TR-EK & PB-LK & PB-RK & PB-EK & MPBK \\
\midrule
Corner Peak & 0.0153/0.0687 & 0.0102/0.0343 & 0.0109/0.0398 & 0.0115/0.0413 &\textbf{0.0095}/0.0343 &0.0104/0.0403 &\underline{0.0096}/0.0317 & 0.0109/0.0038\\
Langermann & 0.1048/0.0359 & \underline{0.1047}/0.0358 & \textbf{0.1033}/0.0355 & 0.1288/0.0781 &0.1048/0.0359	 & 0.3483/0.5793 & 0.2823/0.5881 & 0.1937/0.1877	\\
Rastrigin & 0.6595/0.2543	&\textbf{0.5834}/0.2873	&\underline{0.5945}/0.2520	&0.6441/0.2265	&0.8554/0.7446	&2.4468/1.9668	&3.4097/2.0833 &2.7762/2.4975 \\
Morcaf95a & 0.0309/0.0131 & \textbf{0.01553}/0.0057 & 0.0166/0.0059 & 0.0183/0.0101 & 0.0156/0.0056 & \underline{0.01554}/0.0057 & \underline{0.01554}/0.0057 & 0.0235/0.0123\\
Sphere & 1.4575/0.6646 & 0.7050/0.2009 & \textbf{0.5667}/0.1842 & \underline{0.6957}/0.1773 & 1.4401/1.3483 & 1.5113/1.0441 & 1.6596/1.2412 & 0.9634/0.3520\\
RHE & 1.5364/0.5555 & 1.4885/0.2092 & \underline{1.4450}/0.3984 & \textbf{1.4287}/0.2085 & 1.5208/0.5629 & 1.5722/0.4870 & 1.5556/0.5601 & 1.8379/0.5800\\
Trid & 1.2093/0.0616 & 1.1957/0.0666  & \underline{1.16522}/0.1463 & \textbf{1.16519}/0.1564 & 1.2469/0.0761 & 1.2577/0.0684 & 1.2671/0.0567 & 1.2245/0.1125\\
Schwefel & 0.1938/4.95E-03 & \underline{0.1923}/5.29E-03 & 0.1926/5.53E-03 & \textbf{0.1920}/5.25E-03 & 0.1938/4.95E-03	& 0.1938/4.95E-03	 & 0.1938/4.95E-03  & 0.1927/1.08E-02\\
Stybtang & 0.9185/0.0831 & \underline{0.9175}/0.0825 & \textbf{0.9169}/0.0816 & 0.9176/0.0826 & 0.9185/0.0831 & 0.9185/0.0831  & 0.9185/0.0831 & 0.9300/0.0742\\
\hline
\end{tabular*}
}
\end{table*}

\begin{table*}[h]
\caption{Test results of MAE for 60 training samples (Mean/Std)}\label{tab4}
\resizebox{1\columnwidth}{!}{
\begin{tabular*}{1.45\linewidth}{lllllllll}
\toprule
Functions & UK & TR-LK & TR-RK & TR-EK & PB-LK & PB-RK & PB-EK & MPBK \\
\midrule
Corner Peak & 0.0034/1.22E-03 & 0.0023/8.99E-04 & 0.0023/9.37E-04 & 0.0024/9.58E-04 & \textbf{0.00206}/6.69E-04 &0.0023/1.40E-03	&\underline{0.00214}/7.99E-04  & 0.0024/1.22E-03\\
Langermann & \underline{0.0539}/0.0184 & 0.0544/0.0193 & \textbf{0.0527}/0.0183 & 0.0592/0.0278 &\underline{0.0539}/0.0184	 & 0.1113/0.1261 & 0.0890/0.1224 & 0.0853/0.0744
\\
Rastrigin & 0.2601/0.0780	& \textbf{0.2156}/0.0913	& \underline{0.2294}/0.0755	& 0.2434/0.0781	& 0.3805/0.4131	& 1.0659/0.7429	& 1.3858/0.6969 & 1.2038/0.9947\\
Morcaf95a & 0.0120/0.0056 & \textbf{0.0067}/0.0032 & 0.0071/0.0031 & 0.0074/0.0040 & 0.0068/0.0032 & \underline{0.00676}/0.0032	 & \underline{0.00676}/0.0032 & 0.0084/0.0038
\\
Sphere & 0.8793/0.3633 & 0.4309/0.1314 & \textbf{0.3448}/0.1077 & \underline{0.4240}/0.1086 & 0.8890/0.8174 & 0.9507/0.6850 & 1.0788/0.8230 & 0.5796/0.2058
\\
RHE & 1.1771/0.4455 & 1.1408/0.1707 & \underline{1.1116}/0.3185 & \textbf{1.0958}/0.1790 & 1.1649/0.4489 & 1.2134/0.3850 & 1.1933/0.4469 & 1.4336/0.4804
\\
Trid & 0.9424/0.0533 & 0.9294/0.0576 & \underline{0.9064}/0.1204 & \textbf{0.9050}/0.1324 & 0.9747/0.0618 & 0.9847/0.0575 & 0.9896/0.0419 & 0.9672/0.0980
\\
Schwefel & 0.1553/3.91E-03 & \underline{0.1542}/4.46E-03 & 0.1545/4.68E-03 & \textbf{0.1540}/4.41E-03 & 0.1553/3.91E-03	& 0.1553/3.91E-03	 & 0.1553/3.91E-03 & 0.1545/8.81E-03\\
Stybtang & 0.7346/0.0660 & \underline{0.7338}/0.0655 & \textbf{0.7333}/0.0648 & 0.7339/0.0657 & 0.7346/0.0660 & 0.7346/0.0660  & 0.7346/0.0660 & 0.7445/0.0602\\
\hline
\end{tabular*}
}
\end{table*}

\begin{table*}[h]
\caption{Test results of ${\text{R}^2}$ for 90 training samples (Mean/Std)}\label{tab5}
\resizebox{1\columnwidth}{!}{
\begin{tabular*}{1.48\linewidth}{lllllllll}
\toprule
Functions & UK & TR-LK & TR-RK & TR-EK & PB-LK & PB-RK & PB-EK & MPBK \\
\midrule
Corner Peak & 0.9248/0.0527 & \textbf{0.9441}/0.0489 & 0.9402/0.0440 & 0.9357/0.0519 & 0.9413/0.0374 &\underline{0.9429}/0.0382	&\underline{0.9429}/0.0382  & 0.8974/0.1494\\
Langermann & 0.9953/0.0043 & 0.9982/0.0029 & \textbf{0.99898}/0.00137 & \underline{0.99896}/0.00138 &0.9916/0.0074	 & 0.9740/0.0198 & 0.9875/0.0139 & 0.9849/0.0184\\
Rastrigin & \underline{0.9998}/0.0001	& 0.9922/0.0167	& \textbf{0.9999}/0.0001	& 0.9984/0.0048	& 0.9950/0.0151	& 0.9361/0.0583	& 0.9449/0.0618   & 0.9763/0.0164\\
Morcaf95a & 0.9977/1.33E-03 & 0.99918/4.65E-04 & \textbf{0.9992}/3.28E-04 & 0.99917/5.10E-04 & \underline{0.99919}/4.66E-04 & 0.99918/4.72E-04	 & 0.99917/4.60E-04 & 0.9970/5.82E-03\\
Sphere & 0.9993/0.0007 & 0.99965/0.0003 & \underline{0.99966}/0.0003 & \textbf{0.9997}/0.0002 & 0.9987/0.0027 & 0.9937/0.0059 & 0.9944/0.0054 & 0.9987/0.0020\\
RHE & 0.8646/0.0887 & 0.8738/0.0929 & \textbf{0.8945}/0.1027 & \underline{0.8759}/0.0926 & 0.8716/0.0942 & 0.8719/0.0945 & 0.8557/0.0912 & 0.8025/0.0425\\
Trid & 0.9180/0.0146 & 0.9120/0.0338 & \underline{0.9427}/0.0360 & \textbf{0.9466}/0.0332 & 0.9162/0.0195 & 0.9138/0.0296 & 0.9044/0.0273 & 0.9033/0.0220\\
Schwefel & 0.9842/8.07E-04 & 0.98474/6.30E-04 & 0.9846/8.24E-04 & \underline{0.98479}/5.80E-04 & 0.9842/8.07E-04	& 0.9842/8.05E-04	 & 0.9842/8.04E-04 & \textbf{0.9848}/8.80E-04\\
Stybtang & 0.5403/0.0474 &  0.5412/0.0478 & \underline{0.54130}/0.0477 & 0.54128/0.0478 & 0.5403/0.0474 & 0.5403/0.0474  & 0.5403/0.0474 & \textbf{0.5734/0.0378}\\
\hline
\end{tabular*}
}
\end{table*}

\begin{table*}[h]
\caption{Test results of RMSE for 90 training samples (Mean/Std)}\label{tab6}
\resizebox{1\columnwidth}{!}{
\begin{tabular*}{1.42\linewidth}{lllllllll}
\toprule
Functions & UK & TR-LK & TR-RK & TR-EK & PB-LK & PB-RK & PB-EK & MPBK \\
\midrule
Corner Peak & 0.0109/0.0039 & \textbf{0.0091}/0.0042 & 0.0096/0.0038 & 0.0098/0.0043 & 0.0097/0.0032 & \underline{0.0095}/0.0034	&\underline{0.0095}/0.0034  & 0.0113/0.0071\\
Langermann & 0.0980/0.0508 & 0.0549/0.0414 & \textbf{0.0437}/0.0277 & \underline{0.0446}/0.0272 &0.1307/0.0685	 &0.2423/0.0909 & 0.1544/0.0925 & 0.1672/0.1093	\\
Rastrigin & \underline{0.1373}/0.0500		& 0.4592/0.8194	& \textbf{0.0779}/0.0512	& 0.1996/0.3757	& 0.3390/0.6663	& 2.3421/1.1598	& 1.9360/1.5004 & 1.4135/0.7236\\
Morcaf95a & 0.0233/0.0068 & 0.01418/0.0038 & \textbf{0.0139}/0.0027 & 0.0143/0.0041 & \underline{0.01417}/0.0038 & 0.01419/0.0039	 & 0.0144/0.0037 & 0.0210/0.0195	\\
Sphere & 0.3703/0.2136 & 0.2742/0.1022 & \textbf{0.2615}/0.1324 & \underline{0.2635}/0.0754 & 0.3721/0.4528 & 1.0451/0.7232 & 0.9876/0.6792 & 0.4882/0.3016\\
RHE & 0.9865/0.3514 & 0.9417/0.3693 & \textbf{0.8037}/0.4660 & \underline{0.9335}/0.3657 & 0.9467/0.3821 & 0.9439/0.3850 & 1.0207/0.3557 & 1.2562/0.1425\\
Trid & 0.9872/0.0835 &1.0057/0.2171 & \underline{0.7887}/0.1568 & \textbf{0.7660}/0.2447 & 0.9960/0.1142 & 1.0036/0.1681 & 1.0603/0.0273 & 1.0816/0.1281\\
Schwefel & 0.1906/4.85E-03 & 0.1874/4.36E-03 & 0.1884/5.57E-03 & \underline{0.1871}/4.08E-03 & 0.1906/4.85E-03	& 0.1905/4.86E-03	 & 0.1905/4.86E-03 & \textbf{0.1865}/5.29E-03\\
Stybtang & 0.8413/0.0429 & 0.8404/0.0433 & \underline{0.84037}/0.0433 & 0.84039/0.0433 & 0.8413/0.0429 & 0.8413/0.0429  & 0.8413/0.0429 & \textbf{0.8132}/0.0375\\
\hline
\end{tabular*}
}
\end{table*}

\begin{table*}[h]
\caption{Test results of MAE for 90 training samples (Mean/Std)}\label{tab7}
\resizebox{1\columnwidth}{!}{
\begin{tabular*}{1.46\linewidth}{lllllllll}
\toprule
Functions & UK & TR-LK & TR-RK & TR-EK & PB-LK & PB-RK & PB-EK & MPBK \\
\midrule
Corner Peak & 0.0013/3.68E-04 & \textbf{0.0011}/4.11E-04 & \underline{0.00119}/3.63E-04 & 0.0012/3.24E-04 & 0.00124/2.52E-04 &0.0013/2.42E-04	&0.0013/2.42E-04  & 0.0020/0.0027\\
Langermann & 0.0435/0.0213 & 0.0221/0.0212 & \textbf{0.0133}/0.0057 & \underline{0.0136}/0.0056 &0.0538/0.0246	 & 0.0915/0.0322 & 0.0598/0.0271 & 0.0725/0.0360\\
Rastrigin & \underline{0.0458}/0.0111	 & 0.2322/0.4523	 & \textbf{0.0184}/0.0071	 & 0.0996/0.2553	 & 0.1552/0.3520	 & 1.0863/0.3754	 & 0.8583/0.5892	 & 0.7876/0.3966\\
Morcaf95a & 0.0087/3.08E-03 & 0.00602/1.19E-03 & 0.0061/1.21E-03 & \textbf{0.0058}/9.10E-04 & \underline{0.00600}/1.19E-03 & 0.00603/1.22E-03	 & 0.0062/1.18E-03 & 0.0077/5.49E-03\\
Sphere & 0.2145/0.1105 & 0.1650/0.0560 & \textbf{0.1557}/0.0796 & \underline{0.1586}/0.0449 & 0.2318/0.2939 & 0.6445/0.4407 & 0.6103/0.4116 & 0.2972/0.1964\\
RHE & 0.7432/0.2889 & 0.7011/0.3052 & \textbf{0.5972}/0.3737 & \underline{0.6931}/0.3031 & 0.7105/0.3128 & 0.7073/0.3157 & 0.7724/0.2929 & 0.9625/0.1186\\
Trid & 0.7606/0.0705 & 0.7764/0.1764 & \underline{0.6031}/0.2092 & \textbf{0.5839}/0.1938 & 0.7669/0.0968 & 0.7755/0.1384 & 0.8255/0.1305 & 0.8403/0.1080\\
Schwefel & 0.1527/3.84E-03 & 0.1503/3.57E-03 & 0.1510/4.58E-03 & \textbf{0.1500}/3.40E-03 & 0.1527/3.84E-03	& 0.1527/3.85E-03	 & 0.1526/3.86E-03 & \underline{0.1501}/3.98E-03\\
Stybtang & 0.6728/0.0338 & 0.67214/0.0341 & \underline{0.67210}/0.0341 & 0.67211/0.0341 & 0.6728/0.0338 & 0.6728/0.0338  & 0.6728/0.0338 & \textbf{0.6493}/0.0301\\
\hline
\end{tabular*}
}
\end{table*}

In addition, for global and local accuracy, from Tables \ref{tab3}, \ref{tab4}, \ref{tab6}, and \ref{tab7}, it can be observed that all the TRK models are generally superior to the UK, PBK, and MPBK models. {In fact, the RMSE and MAE are influenced by the nature of the problem. When the magnitude of the target values is large, RMSE and MAE can also be relatively large, even if the model predictions are close to the actual values. This magnitude effect means that, despite accurate predictions, a high value of actual data can result in larger RMSE and MAE. On the other hand, from Tables \ref{tab2}-\ref{tab7}, we can observe that for different PBK models, there are instances where PBK models exhibit the same accuracy and precision as the UK model, and sometimes even perform worse than UK models. This can occur because the PBK models may either not apply regularization to the parameter $\bm{\beta}$ or apply it excessively, which can lead to model degradation.}

The best scenario among all benchmark function tests  is observed in the RHE function test at the initial sample points $n=60$. At this time, the ${\text{R}^2}$ of other Kriging models is less than 0.68 (the worst is 0.6664), while the ${\text{R}^2}$ of all TRK models is greater than 0.72 (the best is 0.7423), which is much better than that of other Kriging models. The RMSE and MAE of the TRK model are much smaller than those of the UK and PBK models. For example, when the initial sample point $n=60$, the RMSE value of approximating the RHE function using the UK model is 1.5364, while the RMSE value of the TR-RK model is only 1.4287. For the standard deviations (Stds) of the ten replicated experiments, it is evident that the Stds of the UK model and the other regularized Kriging models are all consistently small, and the difference is not significant. {Even in most cases, the Stds of the TRK model are better than those of the UK, PBK, and MPBK models.} This result indicates that the proposed TRK models are stable overall, and penalizing the Gaussian stochastic process parameter theta can sometimes improve the stability of the Kriging model.

Different regularized Kriging models offer distinct advantages. Based on the aforementioned results, it is evident that the PBK and MPBK models sometimes perform better in low dimensions. {However, as the dimension of the problem increases, these two models may be over-penalized, resulting in poorer accuracy and stability on the test set, and sometimes performing much worse than the UK model. The reason for this phenomenon is that the PBK and MPBK models essentially penalize the basis function coefficients of the regression process, which may affect the overall prediction trend of the Kriging model.} When the regression process is excessively penalized, the overall performance of the predictive model will become relatively poor. In contrast, the TRK model that penalizes the Gaussian random process parameter $\bm{\theta}$ has almost no significant impact on the overall prediction trend of the Kriging model. In addition, since the Gaussian stochastic process affects the local information of the Kriging model, penalizing the parameter $\bm{\theta}$ of the stochastic process is beneficial to improving the prediction effect of the local area. {Hence, the TRK model exhibits clear advantages over the UK, PBK, and MPBK models.}

Among all Kriging models, TR-RK and TR-EK models generally outperform others in most cases. Therefore, we recommend these models for practical applications, especially the TR-RK model. Because the TR-RK model not only has relatively few parameters but also can effectively improve the accuracy and performance of the Kriging model. Theoretically, if more accurate regularization parameters were chosen, the TR-EK model should significantly outperform the other two TRK models. However, this is not guaranteed. {The reason is that the small cross-validation error in the TR-EK model training set does not necessarily translate to superior prediction performance on the test set.} On the contrary, due to its particularly good performance on the training set, the TR-EK model may exhibit worse results on the test set. {Consequently, the TR-RK model is recommended as the first choice in practical applications.}

The accuracy of the TR-LK model is not particularly outstanding for all tested functions. {However, this does not imply that the method lacks superior predictive performance.} In fact, the TR-LK model may be better suited for addressing high-dimensional problems with strong correlations among features. Therefore, it is also worth considering its application in practical scenarios. In extremely rare cases, the search results for the regularized parameters may not be accurate, resulting in worse outcomes for the TRK model compared to models such as the UK or PBK. Nevertheless, this does not affect the effectiveness of the TRK method, as we only need to provide a sufficiently small interval to search for more accurate penalty coefficients. {Overall, the proposed TRK method is beneficial for improving the precision and accuracy of the model. Consequently, penalizing the Gaussian stochastic process parameter $\bm{\theta}$ of the Kriging model can indeed yield more accurate results and improve the stability of the model.}

\subsection{Engineering examples}\label{subsec6}
{To further validate the accuracy and stability of the TRK model in addressing real-world engineering problems, we conducted numerical simulations using the Borehole \citep{bib51,bib52} and Steel column design optimization \citep{bib53,bib54} problems. Similarly, we compared the results of the TRK models with those of the UK, PBK, and MPBK models.} Except for sampling and verification points, the other parameters and basic settings of the model are consistent with the settings in the test function. All experiments were repeated ten times, and the means and standard deviation of these results were calculated.

\subsubsection{{Borehole simulation}}\label{subsubsec5}
{We first validate the TRK algorithm using the Borehole simulation model. In the field of geological modeling and resource assessment, boreholes are commonly used to obtain information about underground structures, soil properties, groundwater levels, mineral distributions, etc. Sometimes, they are also used to describe the spatial distribution characteristics of underground resources with depth or horizontal direction \citep{bib55}. In slope stability engineering, boreholes are distributed at different locations on the site for better site investigation. Data extracted from boreholes can be used for the stability analysis of random slopes \citep{bib56,bib58} and can be combined with the Kriging method and/or the finite element method to predict soil properties at various depths.}

\begin{figure}[h]
\centering
\includegraphics[width=0.55\textwidth]{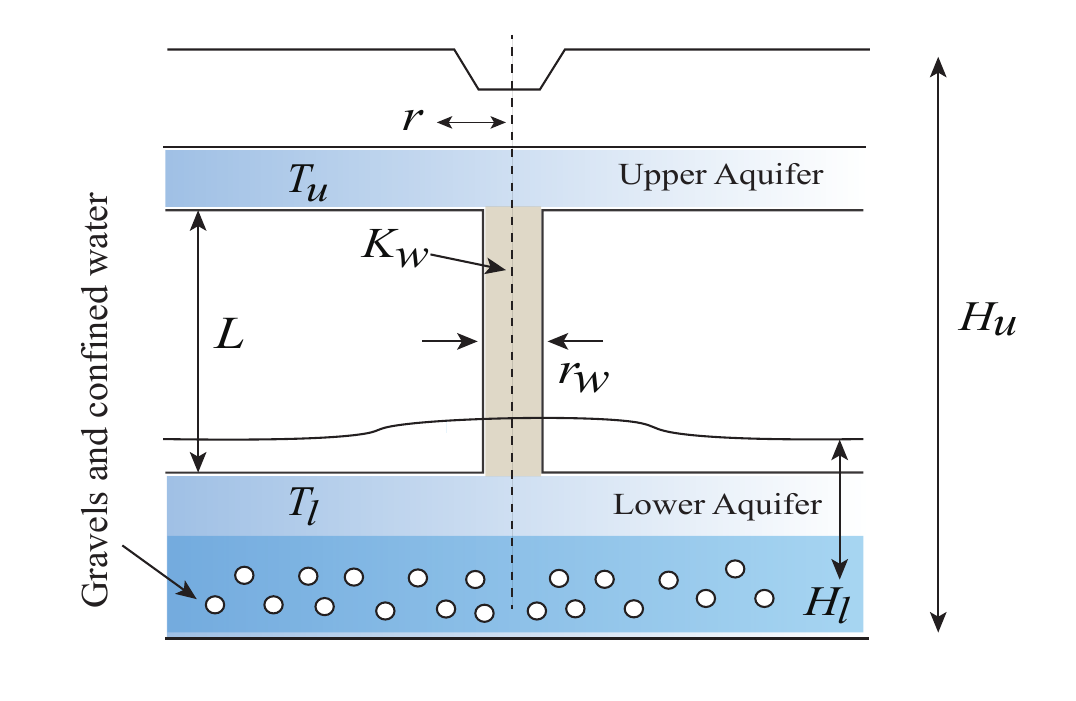}
\caption{Schematic diagram of Borehole simulation}\label{fig5}
\end{figure}

{The simplicity of the Borehole model and its geological background make it widely used for simulation and prediction testing of Kriging modeling to obtain an overall distribution of underground resources. Therefore, it was also considered in this study. This study considers a relatively simple but very common Borehole model \citep{bib51,bib52}, which describes the flow of water through two aquifers from the ground surface, as shown in Fig. \ref{fig5}. The water flow from the ground enters the borehole through the upper aquifer and then flows into the lower aquifer in a steady-state flow. The flow rate of the borehole model is related to eight independent uniformly random variables, including the borehole radius $r_w$, the transmissivity of the upper aquifer $T_u$, the borehole influence radius $r$, the borehole length $L$, and others, as specified in Table \ref{tab8}.}

\begin{table}[h]
\caption{{Specific name and definition domain of Borehole simulation model}} \label{tab8}
\resizebox{0.6\columnwidth}{!}{
\begin{tabular}{lll}
\toprule
Inputs & Specific names of inputs & Domain \\
\midrule
$r_w$    & radius of borehole (m) & $[0.05, 0.15]$ \\
$r$      & radius of influence (m) & $[100, 50000]$ \\
$T_u$    & transmissivity of upper aquifer (m\(^2\)/yr) & $[63070, 115600]$ \\
$H_u$     & potentiometric head of upper aquifer (m) & $[990, 1110]$ \\
$T_l$     & transmissivity of lower aquifer (m\(^2\)/yr) & $[63.1, 116]$ \\
$H_l$     & potentiometric head of lower aquifer (m) & $[700, 820]$ \\
$L$     & length of borehole (m) & $[1120, 1680]$ \\
$K_w$     & hydraulic conductivity of borehole (m/yr) & $[9855, 12045]$ \\ 
\hline
\end{tabular} }
\end{table}

{For geostatistical modeling applications, the flow rate of the borehole model is calculated using the following analytical expression \citep{bib51}:
\begin{equation}
{\mathbf{y}_{h}}=\frac{2\pi {{T}_{u}}({{H}_{u}}-{{H}_{l}})}{\log ({r}/{{{r}_{\omega }}}\;)\left[ 1+\frac{2L{{T}_{u}}}{\log ({r}/{{{r}_{\omega }}}\;)r_{\omega }^{2}{{K}_{\omega }}}+{{{T}_{u}}}/{{{T}_{l}}}\; \right]}, \label{eq49}
\end{equation}}

\begin{figure*}[h]
\centering
\includegraphics[width=1\textwidth]{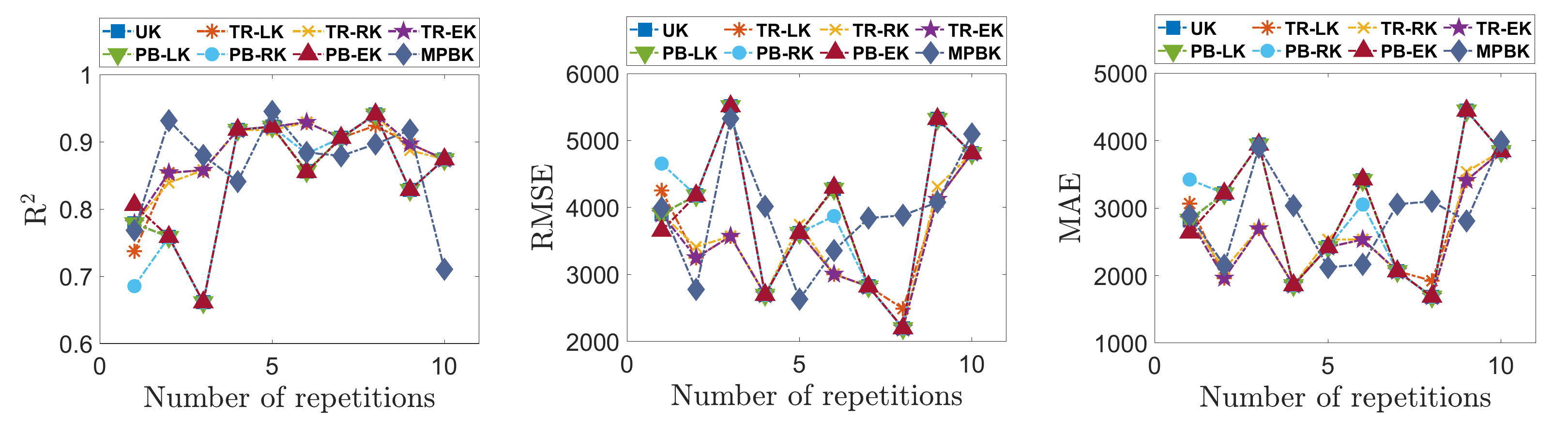}
\caption{{Results of 10 repetitions (Borehole simulation)}}\label{fig6}
\end{figure*}

\begin{figure*}[h]
\centering
\includegraphics[width=1\textwidth]{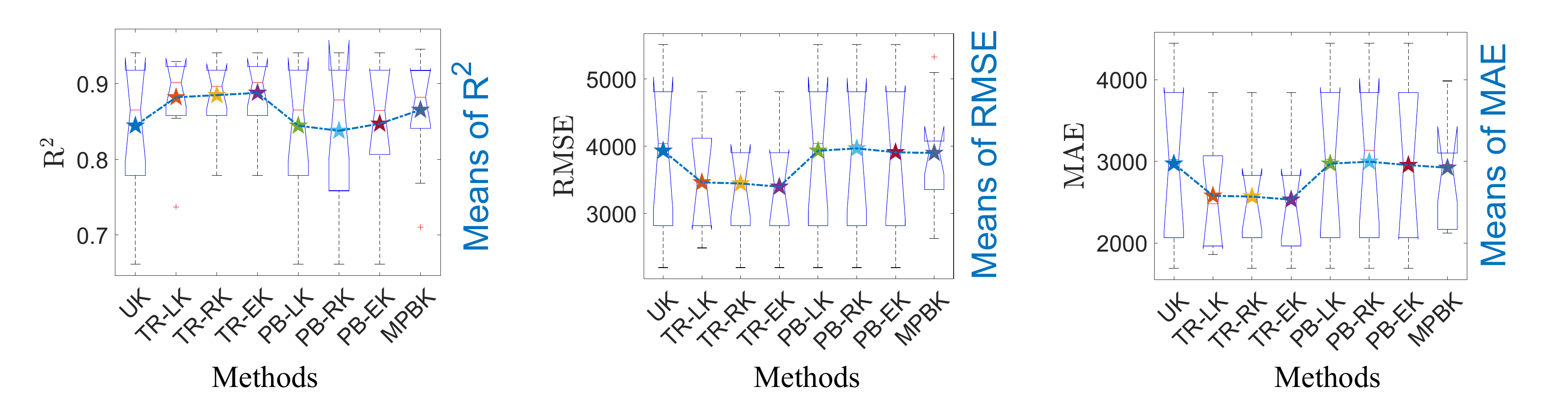}
\caption{{Boxplot of Borehole simulation}}\label{fig7}
\end{figure*}

{Since the flow rate $y_h$ can be expressed as a simple, explicit equation, this model is not computationally complex. However, it is useful for demonstration purposes because its simplicity allows us to quickly assess the accuracy of many test predictions directly. We obtained 80 sample data points from the borehole simulation, which were randomly divided into initial training samples and test samples at a ratio of 3:1. Learning and model construction were performed on the training samples until the stopping criterion was met. Then, the remaining test samples were used to validate the accuracy and stability of the TRK model on this engineering design problem. We compared the results with other regularization Kriging models, and the experimental results are shown in Figures \ref{fig6} and \ref{fig7}. The specific numerical records of the results are listed in Table \ref{tab9}.}

\begin{table}[h]
\caption{{Results of Borehole simulation experiment 8-D (Mean/Std)}} \label{tab9}
\resizebox{1.02\columnwidth}{!}{
\begin{tabular*}{1.64\linewidth}{lllllllll}
\toprule
Indicators & UK & TR-LK & TR-RK & TR-EK & PB-LK & PB-RK & PB-EK & MPBK \\
\midrule
$\text{R}^2$    & 0.8445/0.0889 & 0.8820/0.0579 & \underline{0.8848}/0.0493 & \textbf{0.8879}/0.0485 & 0.8445/0.0889 & 0.8378/0.1016 & 0.8471/0.0870 & 0.8654/0.0740 \\
RMSE   & 3.93e+03/1.12e+03 & 3.46e+03/7.54e+02 & \underline{3.45e+03}/7.91e+02 & \textbf{3.40e+03}/7.67e+02 & 3.93e+03/1.12e+03 & 3.97e+03/1.14e+03 & 3.91e+03/1.12e+03 & 3.90e+03/8.63e+02 \\
MAE    & 2.97e+03/9.52e+02 & 2.58e+03/6.80e+02 & \underline{2.57e+03}/7.00e+02 & \textbf{2.53e+03}/6.93e+02 & 2.97e+03/9.52e+02 & 3.00e+03/9.52e+02 & 2.95e+03/9.58e+02 & 2.92e+03/6.64e+02 \\
\hline
\end{tabular*}
}
\end{table}

{From Fig. \ref{fig6}, in the 10 repeated experiments, the TR-EK and TR-RK models performed relatively well. Specifically, as shown in Fig. \ref{fig7} and Table \ref{tab9}, the TR-EK model has the highest ${\text{R}^2}$ among all Kriging models, with a value of 0.8879. The TR-RK model follows closely with an ${\text{R}^2}$ value of 0.8848. It can be observed that all TRK models perform well in terms of accuracy and stability. The RMSE of the TR-EK model is only 3.40e+03, and the MAE is 2.53e+03, much smaller than those of the UK, PBK, and MPK models. The RMSE and MAE of the TR-RK model are relatively small, with values of 3.45e+03 and 2.57e+03, respectively. In this case, the TR-EK and TR-RK models have certain advantages. Therefore, when applying Kriging models to simulate the Borehole model, it is recommended to prioritize the use of TRK models, especially the TR-EK or TR-RK models.}

\subsubsection{Design optimization problem of steel columns}\label{subsubsec6}
{The second engineering example is the design optimization problem of steel columns \citep{bib53,bib54}. This case involves a nine-dimensional Reliability-Based Design Optimization (RBDO) problem, comprising six environmental variables and three design variables. Balancing the relationship between the cost and reliability of steel columns is a challenging problem, and employing methods such as finite element analysis would consume a significant amount of time and resources. Therefore, surrogate models similar to the Kriging method are often used to simulate this problem \citep{bib59}. In this study, we explored the impact of dependency relationships and different correlation types on the optimization of steel columns under uncertain conditions.}

{The limit state of the steel column is related to the independent uncertain vector $X = (F_s, Z_1, Z_2, Z_3, b, t, h, F_o, E)$ and its stochastic characteristics. The objective of this problem is to minimize the optimal design variables ($b$, $t$, and $h$) under the constraint shown in Eq.(\ref{eq50}), as described in \citep{bib54}. The upper and lower limits of the design variables $(b, t, h)$ are $b = [25, 450]$, $t = [5, 40]$, and $h = [150, 600]$ where $b$ is the average flange breadth, $t$ is the average web thickness, and $h$ is the average profile height, all in millimeters.
\begin{equation}
{{C}_{sc}}~=\left( ~b*t \right)+5*h \label{eq50}
\end{equation}}

{TRK and other Kriging models are constructed as constraints for the RBDO process, as described by the following equation \citep{bib53}:
\begin{equation}
{{\text{g}}_{h}}({{X}_{d}},{{Z}_{d}})={{F}_{s}}-F\left( \frac{1}{2bt}+\left( \frac{{{F}_{0}}}{bth}\cdot \left( \frac{{{\xi }_{b}}}{{{\xi }_{b}}-F} \right) \right) \right) \label{eq51}
\end{equation}
where ${{\xi}_{b}}$ and $F$ represent the Euler buckling load and combined load on the steel column, respectively. $F={{Z}_{1}}+{{Z}_{2}}+{{Z}_{3}}$, ${{\xi}_{b}}=\frac{{{\pi}^{2}}Ebd{{h}^{2}}}{2{{L}^{2}}}$, $L$ is the length of the steel column, with a value of 7500 mm. The detailed information of the nine independent design and environmental variables for this problem is defined as follows: yield stress $F_s$ ($\mu/\sigma = 400/35$ MPa), weight load $Z_1$ ($\mu/\sigma = 500000/50000$ N), variable load $Z_2$ ($\mu/\sigma = 600000/90000$ N), variable load $Z_3$ ($\mu/\sigma= 600000/90000$ N), flange breadth $b$ ($\mu/\sigma= b/3$ mm), flange thickness $t$ ($\mu/\sigma = t/2$ mm), profile height $h$ ($\mu/\sigma = h/5$ mm), initial deflection $F_0$ ($\mu/\sigma = 30/10$ mm), Young’s modulus $E$ ($\mu/\sigma=21000/ 4200$ MPa).} 

\begin{figure*}[h]
\centering
\includegraphics[width=1\textwidth]{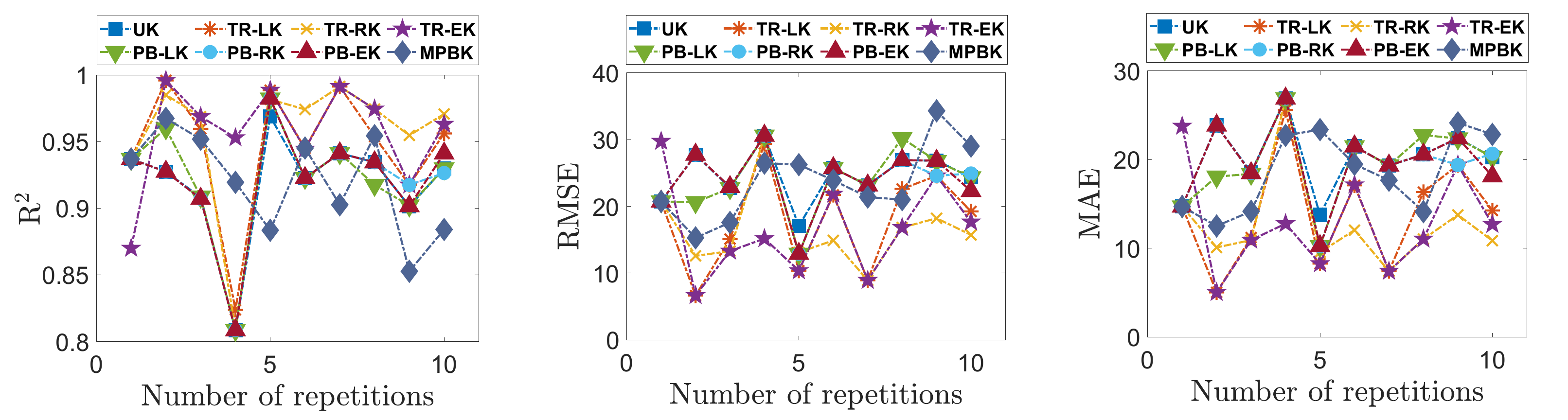}
\caption{{Results of 10 repetitions (Steel columns simulation)}}\label{fig8}
\end{figure*}

\begin{figure*}[h]
\centering
\includegraphics[width=1\textwidth]{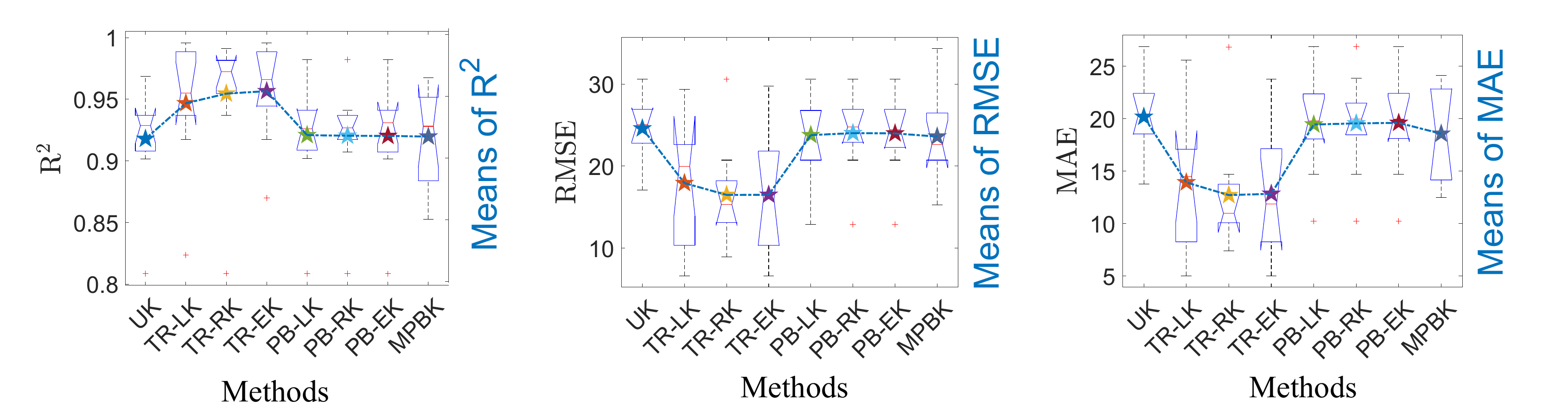}
\caption{{Boxplot of Steel columns simulation}}\label{fig9}
\end{figure*}

{The TRK model was developed following the method outlined in Section \ref{sec3}. We divided 100 simulation samples into initial training and testing sets at a 3:1 ratio. Learning and model construction were performed on 75 training samples until convergence based on the stopping criteria. Subsequently, the remaining 25 testing samples were used to validate the performance of the TRK model and other regularized Kriging models for this engineering design problem. The results are shown in Figures \ref{fig8} and \ref{fig9}, where Fig. \ref{fig8} shows the results of each single experiment and Fig. \ref{fig9} shows the boxplot of the ten experiments. The specific values of these results are recorded in Table \ref{tab10}.}


{From Figs. \ref{fig8} and \ref{fig9}, for ten repetitions of the experiment, the TR-EK and TR-RK models emerged as the top-performing models. Each of these models exhibits distinct advantages: the TR-EK model boasts the highest $\text{R}^2$ value, while the TR-RK model demonstrates smaller RMSE and MAE values. Specifically, as illustrated in Table \ref{tab10}, the TR-EK model achieves an $\text{R}^2$ value of 0.9566, slightly higher than the TR-RK model's 0.9546. However, the TR-RK model outperforms in terms of RMSE and MAE for this problem. The RMSE and MAE of this model are 16.4920 and 12.7193, respectively, while those of the TR-EK model are 16.5092 and 12.8432, respectively. Furthermore, the TR-LK model also produced commendable results. Overall, the prediction accuracy of all TRK models in this scenario is significantly higher than that of the UK, PBK, and MPBK models. In terms of stability, the TRK models also performed relatively well. Therefore, for the steel column design optimization experiment, the TRK model is recommended. In particular, the TR-RK model is emphasized due to its fewer parameters and higher accuracy.}

\begin{table}[h]
\caption{{Results of Steel columns simulation experiment 9-D (Mean/Std)}}\label{tab10}%
\resizebox{1\columnwidth}{!}{
\begin{tabular*}{1.34\linewidth}{lllllllll}
\toprule
Indicators & UK & TR-LK & TR-RK & TR-EK & PB-LK & PB-RK & PB-EK & MPBK \\
\midrule
$\text{R}^2$    & 0.9179/0.0426 & 0.9468/0.0500 & \underline{0.9546}/0.0536 & \textbf{0.9566/0.0387} & 0.9209/0.0461 & 0.9204/0.0441 & 0.9203/0.0450 & 0.9198/0.0377 \\
RMSE   & 24.5915/3.8656 & 17.9302/7.4072 & \textbf{16.4920}/5.9290 & \underline{16.5092}/7.2621 & 23.7774/5.1639 & 24.0171/4.7839 & 23.9825/4.8970 & 23.5979/5.6208 \\
MAE    & 20.1783/3.9386 & 13.9272/6.1736 & \textbf{12.7193}/5.3639 & \underline{12.8432}/5.7621 & 19.4471/4.5912 & 19.5541/4.5877 & 19.6037/4.6963 & 18.5677/4.4956 \\
\hline			
\end{tabular*}
}
\end{table}

{In summary, for the nine numerical functions and two engineering examples, our results indicate that the TRK model outperforms the UK, PBK, and MPBK models in terms of accuracy and stability. For general problems, the recommended model is TR-RK, as it has fewer parameters than or equal to other models, but it performs better in improving model accuracy and demonstrates stronger generalizability.}

\subsection{{Performance of TRK method on special high-dimensional problem}}\label{subsec7}
{Based on the test results in Sections 4.1 and 4.2, it is evident that the TRK algorithm can handle a wide range of problems across various dimensions. Generally, for most high-dimensional problems, the TRK algorithm can maintain high accuracy. Notably, in mid-dimensional engineering cases, the advantages of the TRK algorithm are more pronounced. In this subsection, we will discuss the performance of the TRK algorithm in dealing with high-dimensional complex coupled problems, further illustrating its effectiveness in handling specialized high-dimensional issues, and stating its potential limitations in dealing with such problems.}

{We consider a custom special high-dimensional (SHD) case, where the test case involves the coupling of multiple complex functions, taking the following form:}
\begin{equation}
f\left( \mathbf{x} \right)=\frac{1}{8}\left[\sum\limits_{i=1}^{8}{(\mathbf{x}_{i}^{4}-16\mathbf{x}_{i}^{2}+5{\mathbf{x}_{i}})}\right]+ \frac{1}{40}\left[ \mathop{\sum }_{i=9}^{16}\frac{x_{i}^{2}}{100}-\underset{i=9}{\overset{16}{\mathop \prod }}\,\cos \left( \frac{{{x}_{i}}}{\sqrt{i}} \right)+1 \right] + \frac{1}{100}\left[\mathop{\sum }_{i=\text{17}}^{\text{25}}\mathop{\sum }_{j=\text{1}}^{i}\mathbf{x}_{j}^{2}\right] \label{eqSHD}
\end{equation}
{where ${\mathbf{x}_{i}}\in \left[ \text{0, 1} \right],i=\text{1,...,25}$. It can be observed that this function is relatively complex, and the weight coefficients of the three coupled functions are also decreasing. In this experiment, we chose an initial sample size of 80, with the remaining settings consistent with those in Section \ref{subsec5}. The experiment was repeated ten times, and the average results are shown in Table \ref{tab11}.} 

\begin{table*}[h]
\caption{{Results of Special high-dimensional problem 25-D (Mean/Std)}}\label{tab11}
\resizebox{0.99\columnwidth}{!}{
\begin{tabular*}{1.47\linewidth}{lllllllll}
\toprule
Indicators & UK & TR-LK & TR-RK & TR-EK & PB-LK & PB-RK & PB-EK & MPBK \\
\midrule
$\text{R}^2$ & 0.8324/1.51e-02 & 0.8333/1.52e-02 & \underline{0.83336}/1.52e-02  & \textbf{0.83337}/1.512e-02 & 0.8324/1.512e-02 & 0.8324/1.512e-02 & 0.8324/1.512e-02 & 0.8324/1.51e-02 \\
RMSE & 0.4173/1.64e-02 & 0.4162/1.64e-02 & \underline{0.41618}/1.66e-02 & \textbf{0.41617}/1.65e-02 & 0.4173/1.65e-02 & 0.4173/1.64e-02 & 0.4173/1.64e-02 & 0.4173/1.64e-02 \\
MAE  & 0.3341/1.27e-02 & 0.3332/1.29e-02 & \underline{0.33318}/1.29e-02 & \textbf{0.33317}/1.29e-02 & 0.3341/1.27e-02 & 0.3341/1.27e-02 & 0.3341/1.27e-02 & 0.3341/1.27e-02 \\
\hline			
\end{tabular*}
}
\end{table*}

{From Table \ref{tab11}, it can be observed that for this coupled function, methods like PBK and MPBK that penalize parameter $\bm{\beta}$ seem to fail, whereas the proposed TRK algorithm still shows some improvement. This may be because the TRK algorithm penalizes $\bm{\theta}$, which could be more effective for locally complex functions. Furthermore, we analyzed the $\bm{\theta}$ values after establishing different models, and the results are recorded in Table \ref{tab12}. Due to the high dimensionality of the parameters and the relatively lower weights of the later dimensions, we have only listed the first eight values of the parameters $\bm{\beta}$ and $\bm{\theta}$. It can be observed that the $\bm{\theta}$ values for the UK, PBK, and MPBK models are identical. This outcome suggests that these methods treat each dimension as equally important, which is evidently not reasonable. In contrast, the TRK algorithm applies regularization to $\bm{\theta}$, resulting in seemingly more accurate $\bm{\theta}$ values. The TRK algorithm indicates that the first few dimensions are more significant, aligning better with the actual scenario.} 

\begin{table*}[h]
\caption{{The parameter estimation results of different regularized Kriging models}} \label{tab12}
\resizebox{0.99\columnwidth}{!}{
\begin{tabular*}{1.19\linewidth}{ll}
\toprule
Models & Estimation results of model parameters\\
\midrule
UK &  $\bm{\sigma^2}=\text{4.23e-03},~\bm{\beta}=(\text{-2.79e-16},-0.28,-0.28,-0.36,-0.41,-0.34,-0.31,-0.32
,…),~\bm{\theta}=(10,10,10,10,10,10,10,10,...)$. \\
TR-LK &  $\bm{\sigma^2}=\text{4.23e-03},~\bm{\beta}=(\text{-2.00e-05},-0.28,-0.28,-0.35,-0.41,-0.34,-0.31,-0.32
,…),~\bm{\theta}=(6.38,4.16,2.71,1.77,1.15,0.75,0.65,0.32,...)$. \\
TR-RK &  $\bm{\sigma^2}=\text{4.23e-03},~\bm{\beta}=(\text{-4.34e-06},-0.28,-0.28,-0.35,-0.41,-0.34,-0.31,-0.32,…), ~\bm{\theta}=(6.38,4.16,2.71,1.77,1.15,0.75,0.49,0.32,...)$.\\
TR-EK &  $\bm{\sigma^2}=\text{4.23e-03},~\bm{\beta}=(\text{-4.34e-06},-0.28,-0.28,-0.35,-0.41,-0.34,-0.31,-0.32,…),~\bm{\theta}=(6.38,4.16,2.71,1.77,1.15,0.75,0.49,0.32,...)$. \\
PB-LK &  $\bm{\sigma^2}=\text{4.23e-03},~\bm{\beta}=(\text{-2.79e-16},-0.28,-0.28,-0.36,-0.41,-0.34,-0.31,-0.32,…),~\bm{\theta}=(10,10,10,10,10,10,10,10,...)$.  \\
PB-RK &  $\bm{\sigma^2}=\text{4.23e-03},~\bm{\beta}=(\text{-2.79e-16},-0.28,-0.28,-0.36,-0.41,-0.34,-0.31,-0.32,…),~\bm{\theta}=(10,10,10,10,10,10,10,10,...)$.  \\
PB-EK &  $\bm{\sigma^2}=\text{4.23e-03},~\bm{\beta}=(\text{-2.79e-16},-0.28,-0.28,-0.36,-0.41,-0.34,-0.31,-0.32,…),~\bm{\theta}=(10,10,10,10,10,10,10,10,...)$.  \\
MPBK  &  $\bm{\sigma^2}=\text{4.23e-03},~\bm{\beta}=(\text{-2.79e-16},-0.28,-0.28,-0.36,-0.41,-0.34,-0.31,-0.32,…),~\bm{\theta}=(10,10,10,10,10,10,10,10,...)$.  \\
\hline
\end{tabular*} 
}
\end{table*}

{Additionally, the estimated parameters $\bm{\sigma^2}$ and $\bm{\beta}$ from different models show little numerical difference. This is expected, as these parameters provide the global prediction trend, while $\bm{\theta}$ provides local information. Due to the difficulty in accurately estimating local information for this problem, the TRK algorithm does not exhibit a significant advantage for such high-dimensional, complex coupled functions, although it slightly improves accuracy. Nonetheless, the TRK algorithm remains effective as it retains the original algorithm's capability for accurate global trend prediction while enhancing the local predictive performance of the Kriging model. Consequently, future work could focus on further improving the algorithm to better handle high-dimensional and complex coupled problems.}

\subsection{{Computational complexity and algorithm convergence}}\label{subsec8}
Generally, due to operations such as calculating the inverse of the covariance matrix and performing Cholesky decomposition, the time complexity of the UK model is $\mathcal{O}(n^3)$, where $n$ is the number of sampling points. Compared to the UK model, the TRK, and PBK models require determining regularization parameters, which leads to increased computational time. Although the proposed GSCV algorithm aids researchers in selecting appropriate regularization parameters, this method involves geometric interval search (or geometric grid search) and cross-validation. The time complexity of the TRK and PBK models employing the GSCV method is $\mathcal{O}(C\cdot n^3)$, where $C=k\ast m$, with $k$ representing the cross-validation folds and $m$ indicating the number of partition points in the geometric interval (or geometric grid). This shows that the TRK, PBK, and UK models have the same order of time complexity, although they differ by a constant factor that is influenced by the number of cross-validation folds and the density of geometric interval divisions. Theoretically, if regularization parameters are manually adjusted (or assumed to be given), the runtime for a single run of the UK, PBK, and TRK models would be nearly identical, as no extra time would be necessary to determine the regularization parameters. {For example, Table \ref{tab13} presents the time consumption for building different Kriging models for various test functions and engineering cases, assuming the regularization parameters are provided. The modeling time of the test function shown is the case where the initial point is 60, and all settings are consistent with those described in Sections 4.1 and 4.2. The experiments were repeated ten times, and the Mean CPU time consumed by each algorithm per run was recorded in seconds. The program was executed on a computer with an Intel Core i5-11320H @ 3.20GHz processor and 16 GB of RAM. It is worth noting that the modeling time for each test function is relatively short due to the few initial sampling points and the use of a first-order polynomial function as the regression basis function.}

{From Table \ref{tab13}, we can observe that the modeling times for the TRK model, UK model, and other PBK models are of the same order of magnitude for different test functions. This observation is based on the premise that the regularization parameters for different models have been determined in advance using the GSCV algorithm. Otherwise, the TRK, PBK, and MPBK methods would require significantly more time. Specifically, for different test functions, we can find that as the dimensionality of the problem increases, the required CPU time also increases. For instance, when the test function is 2-D, the modeling time is approximately 4e-3 seconds, whereas, for test functions with dimensions greater than 6-D, the required modeling time increases by an order of magnitude. Overall, we can see that when the regularization parameters are provided, the modeling time for the TRK model is comparable to that of other algorithms, and in some cases, it even shows a certain advantage.}

\begin{table*}[h]
\caption{{Mean CPU time (s) for constructing different Kriging models}}\label{tab13}
\resizebox{0.85\columnwidth}{!}{
\begin{tabular*}{1.04\linewidth}{llllllllll}
\toprule
Functions & Dimensions & UK & TR-LK & TR-RK & TR-EK & PB-LK & PB-RK & PB-EK & MPBK \\ \midrule
Corner Peak & 2 & 4.14e-03 & 3.34e-03 & 3.72e-03 & \underline{3.23e-03} & 3.32e-03 & 3.50e-03 & 3.24e-03 & \textbf{3.12e-03} \\
Langermann & 2 & 3.87e-03 & 3.62e-03 & 3.71e-03 & 3.41e-03 & 3.60e-03 & 3.72e-03 & \underline{3.37e-03} & \textbf{3.32e-03} \\
Rastrigin & 2 & 4.03e-03 & 3.90e-03 & 3.94e-03 & 3.60e-03 & 3.65e-03 & 3.44e-03 & \textbf{3.01e-03} & \underline{3.11e-03} \\
Morcaf95a & 2 & 4.12e-03 & 3.27e-03 & 3.44e-03 & 3.19e-03 & 3.21e-03 & 3.27e-03 & \textbf{3.02e-03} & \underline{3.05e-03} \\
Sphere & 4 & 1.17e-02 & 9.87e-03 & 1.00e-02 & \textbf{9.27e-03} & 1.03e-02 & 1.03e-02 & \underline{9.30e-03} & 1.12e-02 \\
RHE & 6 & 1.22e-02 & 1.12e-02 & 1.16e-02 & \underline{1.07e-02} & 1.10e-02 & 1.09e-02 & \textbf{1.06e-02} & 1.13e-02 \\
Trid & 8 & 1.59e-02 & \underline{1.43e-02} & 1.50e-02 & 1.44e-02 & 1.49e-02 & 1.55e-02 & \textbf{1.39e-02} & 1.46e-02 \\
Borehole & 8 & 1.71e-02 & 1.44e-02 & 1.46e-02 & \textbf{1.39e-02} & 1.54e-02 & 1.57e-02 & 1.48e-02 & \underline{1.43e-02} \\
Steel columns & 9 & 2.95e-02 & 3.05e-02 & 3.15e-02 & 3.10e-02 & 3.13e-02 & 3.04e-02 & \underline{2.89e-02} & \textbf{2.74e-02} \\
Schwefel & 12 & 2.39e-02 & 2.24e-02 & \underline{2.19e-02} & \textbf{2.09e-02} & 2.29e-02 & 2.29e-02 & 2.25e-02 & 2.23e-02 \\
Stybtang & 24 & 6.33e-02 & 4.85e-02 & \underline{4.80e-02} & \textbf{4.69e-02} & 6.28e-02 & 6.26e-02 & 6.21e-02 & 6.33e-02 \\
SHD & 25 & 1.41e-01 & \underline{9.63e-02} & 9.73e-02 & \textbf{9.59e-02} & 1.42e-01 & 1.43e-01 & 1.37e-01 & 1.63e-01 \\ \bottomrule
\end{tabular*}
}
\end{table*}

\begin{figure}[h]
\centering
\includegraphics[width=0.96\textwidth]{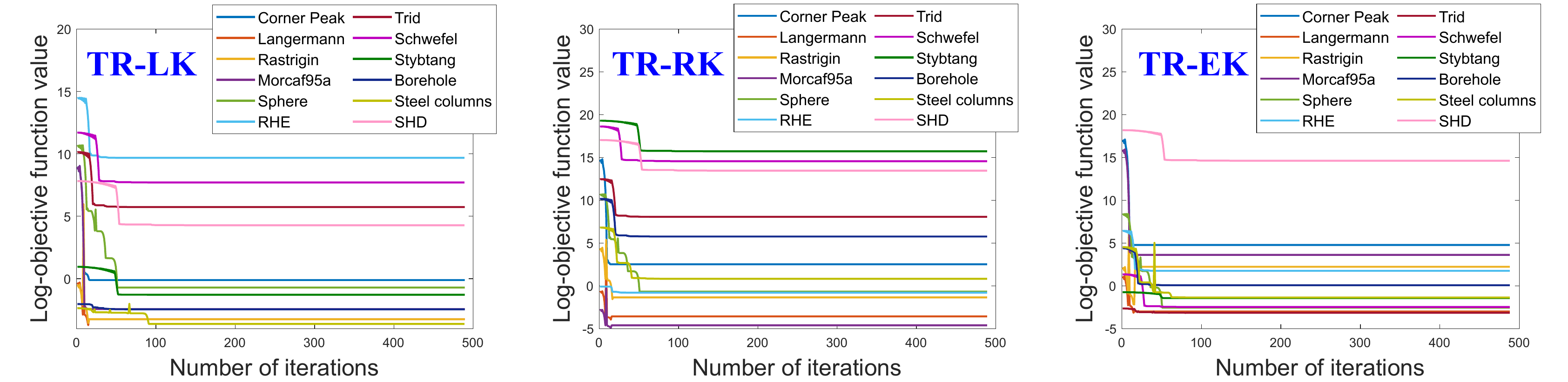} 
\caption{{Iterative curves of TRK models on different test functions}}\label{figiter}
\end{figure}

{In addition, we investigated the convergence behavior of the TRK algorithm. Figure \ref{figiter} presents the convergence curves of the three TRK models for all test and simulation cases. The log-objective function values over 500 iterations are plotted, regardless of whether the termination criteria were met. It can be observed that the log-objective function values of the TRK models consistently decrease, with all three TRK algorithms achieving approximate convergence within 200 iterations for all datasets. These results indicate that the TRK models exhibit good convergence performance.}

\section{Conclusion and discussion}\label{sec5}
In this article, we proposed the TRK models designed to penalize parameters $\bm{\theta}$ of the Gaussian stochastic process. We explored three regularization penalties: $l_1$, $l_2$, and Elastic-net. Compared with the UK, PBK, and MPBK models, the TRK model not only has a significant advantage in finding the minimum of the objective function but also has a chance to obtain a more accurate solution in solving the optimal parameters of the Kriging model. The concrete implementation of the proposed TRK method was provided, including the theoretical derivation, the corresponding optimization iterations, and the model tuning algorithms of the TRK models. The proposed methods were tested using eight numerical functions and two real engineering examples. The results indicate that, in terms of accuracy and stability, all the proposed TRK models outperform the other regularized models. {When dealing with general problems, it is particularly recommended to use the $l_2$ regularization method for constructing the TR-RK surrogate model.} Although TR-EK is also a good choice, it is not recommended in practice, because it has an extra parameter to be tuned. Additionally, due to the stronger penalty of $l_1$ regularization on parameters, the TR-LK method may be the best choice when the data are high-dimensional and have redundant or irrelevant features. Anyway, the proposed TRK model with penalized Gaussian stochastic process parameter $\bm{\theta}$ does improve the accuracy, stability, and generalization ability of the Kriging model. Consequently, the TRK model, especially the TR-RK surrogate, is strongly recommended for practical applications.

The main contributions of this work are as follows:

1) We proposed the TRK model that Penalizes the parameter $\bm{\theta}$ of the Kriging Gaussian stochastic process. This method can more effectively learn the hyperparameters and find more accurate parameter solutions, thereby improving the accuracy and generalization ability of the Kriging model.

2) The Gaussian process parameter $\bm{\theta}$ is penalized from the perspective of the likelihood function when solving the optimal iteration problem of the TRK model. Meanwhile, three regularization methods ($l_1$, $l_2$, and Elastic-net) were considered, and the GSCV methods were combined to obtain the optimal regularization coefficients.

3) The results of the TRK models are interpretable. Meanwhile, a large number of numerical cases were tested to validate the proposed model. In most cases, our results show that the proposed method has better accuracy and stability than the UK and PBK models, especially the TR-RK model.

{The main limitations of the TRK model can be described as:}

{1) The regularization parameters need to be provided by the user. Although we proposed the GSCV algorithm to find the optimal regularization parameters, this method may be computationally inefficient. Particularly, when the search interval is densely divided or the number of cross-validation folds is chosen too large, the time performance of this algorithm can be significantly constrained.}

{2) The essence of the TRK method is to modify the objective function of the Kriging model. Although our experiments show that the method performs well on most problems, its effectiveness may be limited in cases involving multiple singular values or large-scale complex problem couplings due to the inherent limitations of regularization methods and the specific nature of certain problems.}
 
{In conclusion, the proposed TRK algorithm has a certain effect in improving the local prediction performance of Kriging. Future research can develop more efficient regularized parameter search algorithms or further optimize or improve the proposed TRK model.}

\section*{Conflict of interest}
The author declares that there are no conflicts of interest related to the content of this article.

\section*{CRediT authorship contribution statement}
\textbf{Xuelin Xie}: Conceptualization, Formal analysis, Methodology, Software, Visualization, Investigation, Writing - original draft. \textbf{Xiliang Lu}: Conceptualization, Formal analysis, Methodology, Validation, Supervision, Funding acquisition, Writing - review \& editing.

\section*{Acknowledgements}
This work was supported by the National Key Research and Development Program of China (Grant No. 2023YFA1000103) and the National Natural Science Foundation of China (Grant No.12371424). The main numerical calculations have been done at the Supercomputing Center of Wuhan University.

\printcredits

\bibliographystyle{elsarticle-num}
%\bibliographystyle{cas-model2-names}

{\bibliography{Reference}}

%\nolinenumbers

\end{document}